\documentclass[twocolumn,pra,aps,superscriptaddress,showpacs,amsmath,amssymb,floatfix]{revtex4-1}
\usepackage{graphicx}
\usepackage{amssymb}
\usepackage{amsmath}
\usepackage{epsfig}
\usepackage{color}
\usepackage{mathtools}
\usepackage[colorlinks,linkcolor=blue,anchorcolor=blue,citecolor=blue,urlcolor=blue]{hyperref}
\usepackage{physics}
\usepackage{cprotect}
\usepackage{float}
\usepackage{booktabs}
\usepackage{comment}
\usepackage[normalem]{ulem}

\newcommand{\bos}{\mathbf}
\begin{document}


\title{A unified framework for efficient quantum simulation of nonlinear spectroscopy}

\newcommand{\wxy}[1]{\textcolor{red}{Xiaoyang: #1}} 

\author{Long Xiong}
\thanks{These authors contributed equally to the work.}
\affiliation{Center on Frontiers of Computing Studies, Peking University, Beijing 100871, China}
\affiliation{School of Computer Science, Peking University, Beijing 100871, China}
\author{Xiaoyang Wang}
\thanks{These authors contributed equally to the work.}
\affiliation{RIKEN Center for Interdisciplinary Theoretical and Mathematical Sciences (iTHEMS), Wako 351-0198, Japan}
\affiliation{RIKEN Center for Computational Science (R-CCS), Kobe 650-0047, Japan}
\author{Xiaoxia Cai}
\email{xxcai@ihep.ac.cn}
\affiliation{Institute of High Energy Physics, Chinese Academy of Sciences, Beijing 100049, China}
\author{Xiao Yuan}
\email{xiaoyuan@pku.edu.cn}
\affiliation{Center on Frontiers of Computing Studies, Peking University, Beijing 100871, China}
\affiliation{School of Computer Science, Peking University, Beijing 100871, China}

\date{\today }
\begin{abstract}

Nonlinear spectroscopy is a cornerstone of quantum science, providing unique access to multi-point correlations, quantum coherence, and couplings that are invisible to linear methods. However, classical simulation of these phenomena is fundamentally limited by the exponential growth of the Hilbert space, and practical quantum algorithms for the nonlinear regime have remained largely unexplored. Here, we present a unified quantum algorithmic framework for computing $n$-th order nonlinear spectroscopies. By reformulating multi-time responses as a weighted sum of expectation values at finite pump amplitudes via a generalized parameter shift rule, our approach bypasses the costly evaluation of high-order commutators and time-dependent operator expansions. This reformulation enables efficient execution via real-time evolution on current quantum hardware, ensuring inherent noise resilience. We validate the framework on IBM's superconducting quantum processors, successfully obtain higher-order response functions of a 12-qubit XXZ spin-chain. Furthermore, the versatility of our method is demonstrated by resolving quasi-particle excitation spectra in spin-liquids and identifying interaction-induced cross-peaks in atomic systems. Our results establish a practical and scalable pathway for probing complex quantum dynamics on near-term quantum devices, extending the reach of quantum simulation into the nonlinear domain.
\end{abstract}

\maketitle

\begin{figure*}
	\centering
	\includegraphics[width=0.85\textwidth]{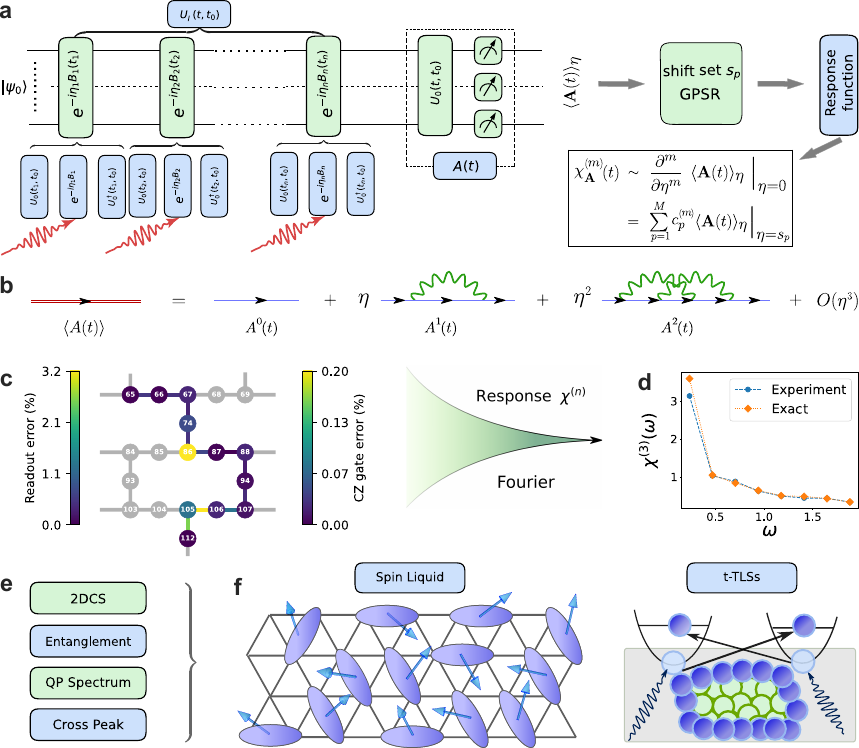}
\caption{Schematic of the quantum algorithm for nonlinear response extraction and its applications. (a) Quantum-circuit schematic for simulating spin-lattice dynamics driven by a sequence of impulsive pumps. The upper row shows the interaction-picture representation, where the driven evolution is written as a sequence of time-local kicks $e^{-i\eta_j B_j(t_j)}$ forming $U_I(t,t_0)$, while the lower row illustrates the equivalent Schr\"odinger-picture factorization as free evolutions $U_0(t_{j+1},t_j)$ interleaved with instantaneous kicks $e^{-i\eta_j B_j}$. Measuring the probe yields the time-resolved observables $\langle \mathbf A(t)\rangle_{\vec\eta}$. The values measured at a finite shift set $\{s_p\}$ are combined via the GPSR to reconstruct the target response function. (b) The measured signal is expanded in powers of the perturbation strength $\vec\eta$, with each coefficient corresponding to a distinct response order to be isolated. (c) Experimental context (left) and spectral analysis (right): the selected qubit subgraph on the device, and the Fourier transform from $\chi^{(n)}(t)$ to $\chi^{(n)}(\omega)$. (d) Example of a third-order frequency-domain response $\chi^{(3)}(\omega)$, showing close agreement between QPU data and exact simulation. (e) Applications enabled by our method, including 2D coherent spectroscopy (2DCS), entanglement dynamics, quasi-particle (QP) excitation spectra, and cross-peaks. (f) Representative target platforms for these applications: a spin-liquid model (left) and interacting two-level systems (right).}
	\label{fig1}
\end{figure*}

Understanding the response of quantum systems to external perturbations is central to modern science and forms the theoretical foundation of experimental probes such as Raman scattering, pump-probe spectroscopy, dynamical structure factors, and nuclear magnetic resonance~\cite{molspectro1,molspectro2,pumpprobe,dynamicstruc,nakatsukasa2005linear,buraczynski2016static,o2015predicting,boschini2024time}. Many phenomena of fundamental and practical interest lie beyond the linear-response regime~\cite{Mukamel1995,RevModPhys.88.045008}, where nonlinear response theory provides a unified description and underpins nonlinear and multidimensional spectroscopies, including two-dimensional coherent and optical techniques, that have yielded fundamental insights into quasiparticle dynamics, energy transfer, and entanglement in quantum materials and molecular systems~\cite{jansen2019theory,noda2004advances,cho2019introduction,wu2018two,li2021photon,mcginley2024signatures,kanega2021linear,choi2020theory,knuppel2019nonlinear,xu2024hexadecapole,maly2023separating,ohno2016phase,inoue2020reorientation,ohno2017second,meinel2022quantum,trovatello2022disentangling,cai2020quantum,hata2021three,liebel2018room,yang2024phantom}. Yet, classical simulation of such dynamical responses remains challenging due to the exponential growth of the Hilbert space with system size.

Quantum computing offers a transformative alternative to these classical limitations. Specifically, quantum simulation~\cite{feynman2018simulating,lloyd1996universal} has emerged as a promising approach to accessing many-body dynamics beyond the reach of classical computers~\cite{harrow2017quantum,preskill2018quantum,daley2022practical,nielsen2010quantum}. In recent years, substantial progress has been made in measuring linear-response properties using quantum simulators, employing strategies such as direct frequency-domain evaluation and extracting response functions from the lowest-order commutator~\cite{kokcu2024linear,2025linearresponse,wang2025computing,huang2022variational,cai2020quantum,kosugi2020linear,2019np,2022prr,2021pra,2020prbfrancis,PhysRevD.102.094505}. Beyond linear response, the computation of nonlinear response using quantum computers involves evaluating high-order commutators. Traditional methods express these quantities as linear combinations of many multi-time correlation functions, or require computationally demanding ancillary qubits~\cite{kharazi2025efficient,loaiza2025nonlinear}, hindering the implementation of these algorithms on currently available quantum devices.

In this work, we present an efficient quantum algorithmic framework for computing nonlinear response functions of many-body systems (see Fig.~\ref{fig1}). This framework uses the observation that the nonlinear response function is a higher-order derivative of the response expectation values, and the higher-order derivative can be efficiently evaluated using the generalized parameter-shift rule (GPSR)~\cite{PhysRevA.98.032309,PhysRevA.99.032331,crooks2019gradients,wierichs2022general,abramavicius2025evaluationderivativesusingapproximate}. In this way, we reformulate the nonlinear response function as a linear combination of expectation values evaluated at finitely many time-ordered pump amplitudes, and the expectation values can be efficiently obtained through real-time evolution on quantum simulators. Thus, this approach avoids the need to compute many correlation functions or introduce ancillary qubits, reduces experimental overhead, and is therefore better suited for noisy and early fault-tolerant quantum simulators.

The approach is benchmarked on Heisenberg XXZ model and experimentally validated on IBM's superconducting quantum processors using 12 qubits~\cite{cross2017open,cross2022openqasm,qiskit}. We evaluate the model's fourth-order responses of two-site magnetization and spin current, as well as the fifth-order spin responses. Using error mitigation methods, the measured nonlinear response functions are consistent with the noiseless results in both time and frequency domains. We further demonstrate the versatility of the method through applications to spin-liquid and atomic systems, capturing quasi-particle excitation spectra and interaction-induced cross-peaks. Our approach establishes a scalable framework for computing nonlinear responses and multi-dimensional spectroscopy in the presence of realistic quantum noise.

\noindent\textbf{Results}\\
\noindent\textbf{Framework for nonlinear responses}\\
First, we briefly outline the formalism of higher-order multi-time nonlinear response functions formulated in the interaction picture, which forms the standard theoretical description of nonlinear spectroscopy.
Consider a physical system with unperturbed Hamiltonian $\mathbf H_0$, driven by external classical fields through an interaction Hamiltonian $\mathbf H_I(t)$. The density operator $\boldsymbol{\rho}$ evolves according to the Liouville–von Neumann equation $i \partial_t \boldsymbol{\rho}(t) = [\mathbf H_I(t), \boldsymbol{\rho}(t)]$, where the coupling to classical source fields is modeled by
\begin{align}
    \mathbf H_I(t) = \sum_{a=0}^{L-1} h_a(t )\, \mathbf B_a(t).
    \label{eq:external-field}
\end{align}
Here, $h_a(t)$ and $\mathbf B_a(t)=e^{i\mathbf H_0 t}\mathbf B_a e^{-i\mathbf H_0 t}$ denote the external field and the system operator coupled to the external field, and the index $a$ labels independent coupling channels (e.g., Cartesian components, polarizations, internal degrees of freedom).

For a system initially prepared in
$|\psi_0\rangle$, the expectation value of a physical observable $\mathbf{A}$ in the presence of the external fields $\vec h=(h_0,h_1,\cdots,h_{L-1})$ measured at time $t$ reads
\begin{equation}
\langle \mathbf A(t) \rangle_{\vec h}
=\bra{\psi_0} U_I^\dagger(t,-\infty)\,\mathbf A(t)\,U_I(t,-\infty) \ket{\psi_0},
\end{equation}
with the time propagator
$U_I(t,-\infty)=\mathcal{T} e^{-i \int_{-\infty}^{t} H_I(t') dt'}$. Next, we expand $\langle\mathbf A(t)\rangle_{\vec h}$ in powers of the source fields as
\begin{align}
    \langle \mathbf A(t)\rangle_{\vec h}=\sum_{m=0}^{\infty} A^m(t),
    \label{eq:Am}
\end{align}
with the unperturbed term $A^0(t) = \langle \mathbf A(t)\rangle_{\vec h = \vec 0}$ and 
\begin{equation}\label{eq:Am_sums}
    \begin{aligned}
A^m(t)
=
&\sum_{a_1,\dots,a_m=0}^{L-1}
\int dt_1\cdots\int dt_m\\ 
& \chi^{(m)}_{\mathbf A;a_1\cdots a_m}(t;t_1,\dots,t_m)\;
\prod_{k=1}^m h_{a_k}(t_k),
    \end{aligned}
\end{equation}
    where $\chi^{(m)}_{\mathbf A;a_1\cdots a_m}(t;t_1,\cdots,t_m)=\left.
    \frac{\delta^m\langle\mathbf A(t)\rangle_{\vec h}}
    {\delta h_{a_m}(t_m)\cdots\delta h_{a_1}(t_1)}
    \right|_{\vec h=\vec0}$ is the $m$th-order $(m+1)$-time response kernel (susceptibility), and can be equivalently expressed via the Kubo-type nested-commutator representation~\cite{Mukamel1995}:
\begin{align}
    & \chi^{(m)}_{\mathbf A;a_1\cdots a_m}(t;t_1,\dots,t_m)
     = i^m
       \theta(t-t_1)\cdots\theta(t_{m-1}-t_m) \nonumber \\
    &\qquad\qquad \big\langle\psi_0\big|
    \operatorname{ad}_{\mathbf B_{a_m}(t_m)}\cdots\operatorname{ad}_{\mathbf B_{a_1}(t_1)}
    \mathbf A(t) \big|\psi_0\big\rangle.
    \label{eq:nth-response-function}
\end{align}
Here, $\operatorname{ad}_{\mathbf X}(\mathbf Y):=[\mathbf X,\mathbf Y]$ and $\mathbf B_{a}(t)=e^{i\mathbf H_0 t}\,\mathbf B_{a}\,e^{-i\mathbf H_0 t}$.
For $m=1$ and a single channel, it reduces to the standard Kubo formula
$\chi^{(1)}(t;t_1)=i \theta(t-t_1) \bra{\psi_0}[\mathbf B(t_1),\mathbf A(t)]\ket{\psi_0}$ \cite{kubo1957statistical,kokcu2024linear,hertel2004lectures}.

A direct approach of evaluating Eq.~\eqref{eq:nth-response-function} expands its nested commutators into many operator strings and estimate each resulting multi-time correlator using Hadamard-test circuits~\cite{PhysRevLett.113.020505}. Such implementations typically require ancilla qubits and controlled operations, leading to deep circuits that are costly on current hardware with finite qubit connectivity.




Instead, the response coefficient can be evaluated without ancilla qubits and controlled operations by writing the response coefficient as mixed derivatives of $\langle \mathbf A(t)\rangle_{\vec h}$. For simplicity, we model the external driving as a finite sequence of short pulses applied at prescribed times, which is frequently realized in nonlinear spectroscopy and pump-probe experiments. In this setting, the source field $h_a(t)$ in Eq.~\eqref{eq:external-field} is a multi-channel $\delta$-pump: 
\begin{align}
    h_a(t)=\eta_a\sum_{b=0}^{n_a-1}\delta(t-t_{ab}),
\end{align}
i.e., in each channel $a$, the $b$th pulse ($b=0,1,\ldots,n_a-1$) occurs at time $t_{ab}$, sharing the same amplitude $\eta_a$. In the following content, we denote $\vec h=(h_0, \dots, h_{L-1})$, and $\vec \eta=(\eta_0,\dots,\eta_{L-1})$. 

For a single channel in Eq.~\eqref{eq:Am_sums} specified by a sequence $(a_1,\ldots,a_m)$ with $a_i\in\{0,\ldots,L-1\}$, we define the counting vector $\vec\beta:=(\beta_0,...,\beta_{L-1})$ by $\beta_k:=|\{i\in[m]:a_i=k\}|$ with $[m]:=\{1,\ldots,m\}$, and the support $\mathrm{supp}(\vec{\beta}):=\{a:\beta_a>0\}$. Then the $m$th-order response coefficient to the multi-channel $\delta$-pump can be expressed by mixed derivatives of $\langle \mathbf A(t)\rangle_{\vec\eta}$


\begin{align}
	&\sum_{\substack{1 \le b_k \le n_{a_k-1} \\ k=1,\ldots,m}}\chi^{(m)}_{\mathbf A;a_1\cdots a_m}\big(t;\,t_{a_1 b_1},\dots,t_{a_m b_m}\big) \nonumber \\
    = & \frac{1}{\Big(\prod_{k=0}^{L-1}\beta_k!\Big)} \left.\partial_{\vec\eta}^{\,\vec\beta}\,\langle\mathbf{A}(t)\rangle_{\vec \eta}\right|_{\vec\eta=0},
	\label{eq:derivative-to-commutator}
\end{align}
where $\partial_{\vec\eta}^{\,\vec\beta}
	={\partial^{m}}/{\partial \eta_0^{\beta_0}\cdots\partial \eta_{L-1}^{\beta_{L-1}}}$. This mixed derivatives can be numerically evaluated using finite difference method. However, the truncation error from a small finite difference is hard to control, and noise in quantum circuits estimating $\langle\mathbf{A}(t)\rangle_{\vec \eta}$ is inevitably amplified, especially for highly nonlinear spectroscopy with a large $m$~\cite{kokcu2024linear}. In the following content, we introduce the multi-parameter GPSR to evaluate the mixed derivatives without the truncation error.

\begin{figure*}
	\centering
	\includegraphics[width=0.95\textwidth]{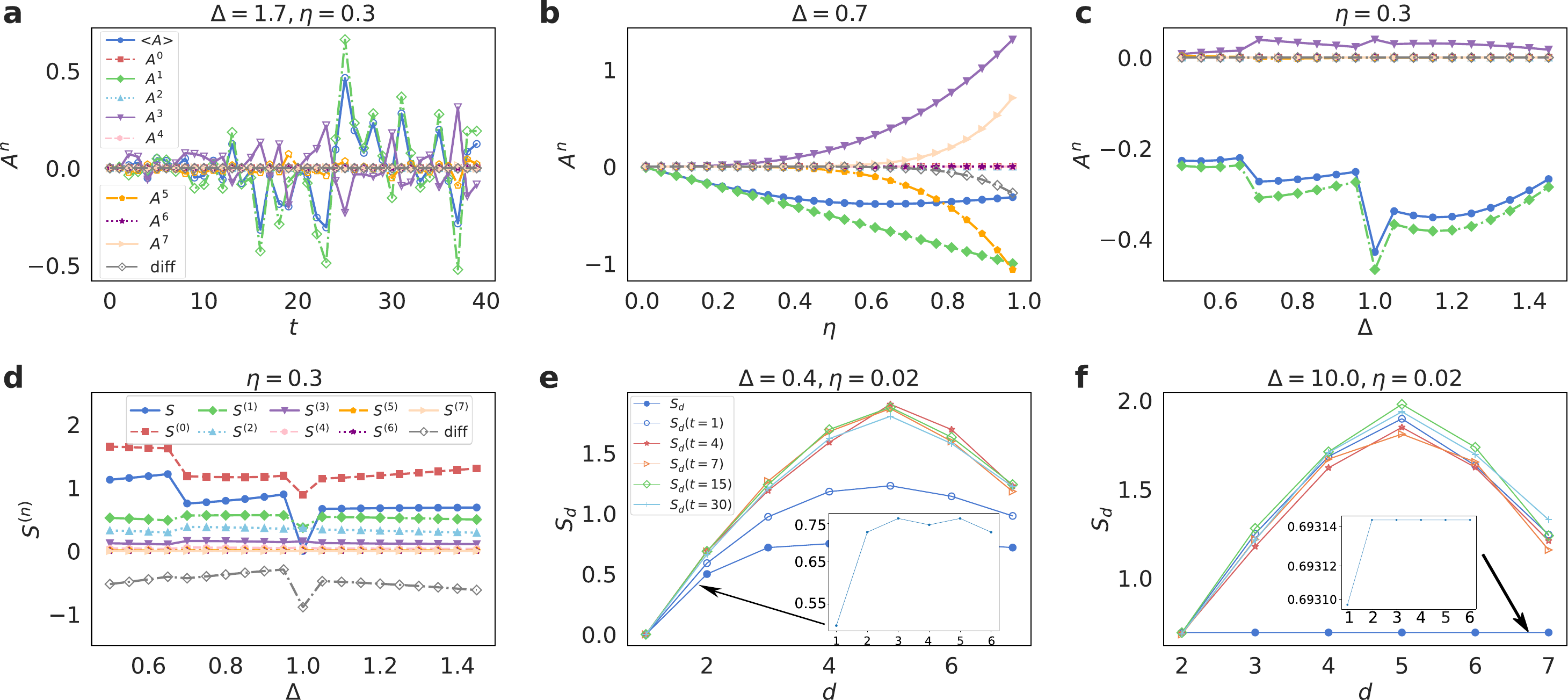}
	\caption{Numerical simulation results in the XXZ model validating the effectiveness of the GPSR method. (a) Reconstructed response contributions $\{A^n(t)\}$ (up to $n=7$) together with the reference signal $\langle \mathbf A(t)\rangle_{\eta}$; the partial sum $\sum_{n=0}^{7}A^n(t)$ accurately approximates $\langle \mathbf A(t)\rangle_{\eta}$ in the perturbative regime. (b) As the single-pump amplitude $\eta$ increases, higher-order terms become more important: the truncation residual $\mathrm{diff}(t)=\langle \mathbf A(t)\rangle_{\eta}-\sum_{n=0}^{7}A^n(t)$ grows with $\eta$, indicating that orders beyond seven become relevant at stronger pumping. (c)–(d) $A^n$ and the corresponding $S^{(n)}$ versus $\Delta$, where $S^{(n)}$ denotes the order-$\eta^n$ term in the entropy expansion; a sharp feature appears near the phase boundary \cite{laflorencie2016quantum}. (e)–(f) Subsystem entanglement entropy $S_d(t)$ for a contiguous block of $d$ sites at multiple times $t$ after the $\delta$-pulse. The curve labeled $S_d$ is the unperturbed ($t=0$) ground-state entropy, whereas $S_d(t\neq 0)$ corresponds to the driven state at later times. Panels (e) and (f) show representative gapless and gapped regimes (e.g., $\Delta=0.4$ and $\Delta=10.0$ at $\eta=0.02$); insets zoom in on the $t=0$ scaling.}
	\label{fig2}
\end{figure*}

\vspace{0.2cm}
\noindent\textbf{Quantum algorithm for nonlinear spectroscopy}



\noindent The multi-parameter GPSR evaluates the mixed derivative in Eq.~\eqref{eq:derivative-to-commutator} by assuming that the pump generator $\bos{B}_a$ admits simple eigenvalues $\{\lambda_{a,s}\}_{s\in [S]}$, and we introduce the set of distinct positive differences $\Omega_a:=\{\lambda_{a,s}-\lambda_{a,s'}|s,s'\in[S], \lambda_{a,s}>\lambda_{a,s'}\}$. Using these eigenvalue differences, the expectation $\langle\mathbf{A}(t)\rangle_{\vec \eta}$ can be written as a finite discrete Fourier series of $\vec{\eta}$ with $|\Omega_a|$ frequency components. Then the mixed derivatives in Eq.~\eqref{eq:derivative-to-commutator} is a linear combination of a multi-dimensional Fourier coefficients of $\langle\mathbf{A}(t)\rangle_{\vec \eta}$, and thus is a linear combination of $\langle\mathbf{A}(t)\rangle_{\vec \eta}$ by the reverse discrete Fourier transformation. In the $a$th dimension of the Fourier series, the number of Fourier coefficients equals to $|\Omega_a|$, and is determined by evaluating $\langle\mathbf{A}(t)\rangle_{\vec \eta}$ at $|\Omega_a|$ shift points $\{s_{a,p_a}\}, p_a\in[|\Omega_a|]$. The detailed derivation is given in Supplemental Material.

Therefore, the $m$th-order nonlinear response coefficient can be written as a linear combination of $\langle\mathbf{A}(t)\rangle_{\vec \eta}$
\begin{equation}
	\begin{aligned}	     
    &\sum_{\substack{1 \le b_k \le n_{a_k-1} \\ k=1,\ldots,m}}
		\chi^{(m)}_{\mathbf A;a_1\cdots a_m}\big(t;\,t_{a_1 b_1},\dots,t_{a_m b_m}\big) \\
		=&\frac{1}{(\prod_{k=0}^{L-1}\beta_k!)}\sum_{\substack{1 \le p_a \le |\Omega_a| \\ a=0,\ldots,L-1}}
		\Bigg[\prod_{a=0}^{L-1} c_{a,p_a}^{(\beta_a)}\Bigg]\,
		\langle\mathbf A(t)\rangle_{\vec\eta= \vec{s}},
	\end{aligned}
\end{equation}
where $\vec{s}=( s_{0,p_0},\dots,s_{L-1,p_{L-1}})$, and $c_{a,p_a}^{(\beta_a)}$ are the GPSR weights for $\mathbf B_a$ at order $\beta_a$. These weights are determined by classically solving a linear system of equations
\begin{align}
\sum_{p=1}^{|\Omega_a|} e^{i\omega s_{a,p}} c_{a,p}^{(\beta_a)} =(i\omega)^{\beta_a},\qquad \forall\,\omega\in\Omega_a,
\end{align}
where $e^{i\omega s_{a,p}}$ is an element of an $|\Omega_a|\times |\Omega_a|$ Fourier transformation matrix. By choosing appropriate shift points, this matrix has a nonzero determinant, and hence the linear system admits a unique solution. In this way, the response coefficient is reconstructed as a finite weighted sum of circuit measurements of $\langle\mathbf A(t)\rangle_{\vec\eta}$ on the corresponding shift grid.

The GPSR overhead is determined by the number of distinct differences $|\Omega_a|$ of the pump generator $\mathbf B_a$, since exact derivative reconstruction requires at least $|\Omega_a|$ shift points. On the other hand, if the derivative is not taken for a direction $a$, i.e., $\beta_a=0$, the GPSR at this direction is not required. Thus, the total number of quantum circuits to evaluate the nonlinear spectroscopy at one time point reads
\begin{align}
    N_{\rm circ}=\prod_{a\in\mathrm{supp}(\vec{\beta})} |\Omega_a|.
    \label{eq:N-circuits}
\end{align}
This circuit complexity has no dependence on the non-linear order $m$, which is distinguished from the conventional methods using Hadamard test and finite difference, making the GPSR method suitable to evaluate highly nonlinear spectroscopy. 

As per Eq.~\eqref{eq:N-circuits}, the efficiency of the GPSR method is guaranteed if all $|\Omega_a|,~a\in\mathrm{supp}(\vec{\beta})$ grow at most polynomially with system size, which is the case for spatially local operators, few-body observables, and simple collective drives such as the zero-momentum mode or a small set of commensurate momentum components with polynomially bounded distinct gap sets. These operators cover the physically relevant pumps commonly used in spectroscopy, where the external field typically couples to dipole/polarization, current-like, or magnetization/spin-like collective observables~\cite{Mukamel1995,Abramaviius2009CoherentMO,lu2017coherent,PhysRevB.93.195144}, rather than to a generic many-body operator resolving exponentially many microscopic configurations. More broadly, conventional simulation methods are generally susceptible to the rapid growth of Hilbert-space complexity, and classical methods---including exact diagonalization, tensor-network~\cite{PhysRevLett.124.137701,PhysRevLett.133.026401,RevModPhys.77.259}, and structured open-system solvers~\cite{PhysRevLett.129.230601,Strathearn2017EfficientNQ}---typically rely on favorable conditions such as restricted system size, controlled entanglement growth, or finite bath-memory/mode structure. By contrast, the dominant overhead of the GPSR method is governed by the spectral structure of the chosen pump operators rather than by the full Hilbert-space dimension. 

\begin{figure*}[!htbp]
	\centering
	\includegraphics[width=0.9\textwidth]{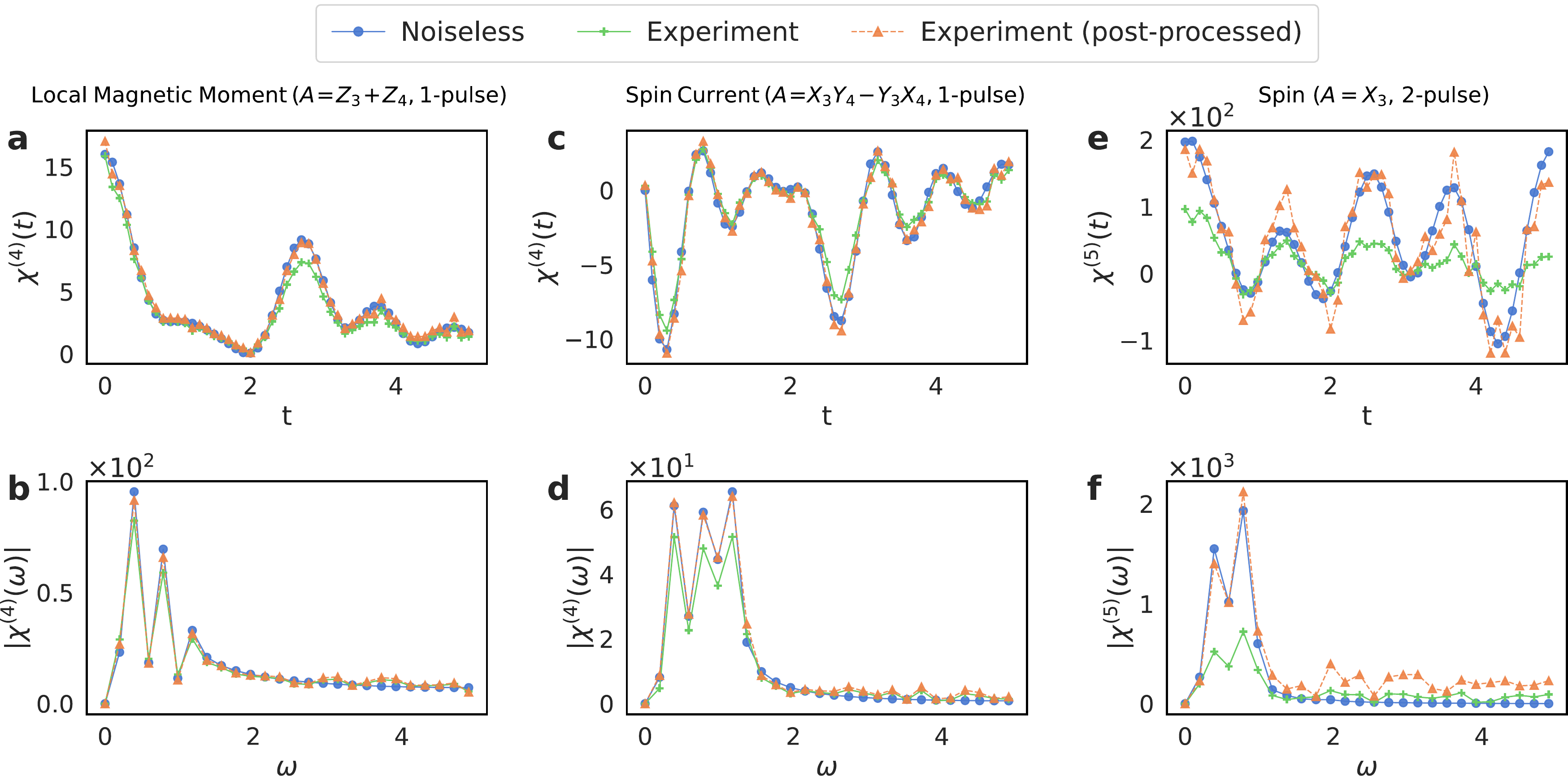}
    \cprotect\caption{Experimental implementation and verification of our quantum algorithm. We implement the XXZ spin-chain dynamics on superconducting quantum processor \verb|ibm_kobe|. Panels (a)–(f) compare fourth- and fifth-order nonlinear response functions in the time and frequency domains for different observables in a 12-qubit XXZ chain ($N=12$ with $\Delta = 0$, $h_e = 0.75$), reconstructed from hardware measurements at shifted pump amplitudes. The evolution time is sampled over $t\in[0,5]$ with 51 points. In (a), we show the time-domain fourth-order response $\chi^{(4)}(t)$ of the two-site magnetization, while (b) displays the corresponding spectrum $|\chi^{(4)}(\omega)|$ obtained via real FFT after mean subtraction. In (c), the time-domain fourth-order response $\chi^{(4)}(t)$ of the spin current, and (d) gives the corresponding spectrum $|\chi^{(4)}(\omega)|$. Panel (e) presents the time-domain fifth-order response $\chi^{(5)}(t)$ for $\mathbf A = X_3$ under a two-pulse protocol with identical pumps $\mathbf B = X_3$ at $t=0$ and $t=1$, with the associated frequency spectrum $|\chi^{(5)}(\omega)|$ shown in (f). Blue lines with circles denote noiseless simulations, orange dashed lines with triangles denote post-processed experimental data obtained via an empirical exponential envelope fit $a(t)=\alpha e^{-\gamma t}$, and green solid lines with plus markers represent the raw experimental data from the quantum processor.}
	\label{fig3}
\end{figure*}

As an example, consider a single pumped channel $a$ with one $\delta$-pulse at $t_{a0}$ (i.e., $n_a=1$ and $h_a(t)=\eta\,\delta(t-t_{a0})$). The $m$th-order response in Eq.~\eqref{eq:derivative-to-commutator}  reduces to
\begin{align}
	\chi^{(m)}_{\mathbf A;\underbrace{a\cdots a}_{m}}\big(t;\underbrace{t_{a0},\dots,t_{a0}}_{m}\big)
	=
	\frac{1}{m!}\left.\frac{\partial^m}{\partial\eta^m}\langle\mathbf A(t)\rangle_{\eta}\right|_{\eta=0},
\end{align}
where $\langle\mathbf A(t)\rangle_{\eta}$ is the circuit-measured expectation at pulse amplitude $\eta$. The derivative is then obtained exactly via the single-parameter GPSR,
\begin{align}
	\left.\frac{\partial^m}{\partial\eta^m}\langle\mathbf A(t)\rangle_{\eta}\right|_{\eta=0}
	=
	\sum_{p=1}^{M}c_{p}^{(m)}\left.\langle\mathbf A(t)\rangle_{\eta}\right|_{\eta=s_{p}},
\end{align}
with $\{s_p\}$ and $\{c_p^{(m)}\}$ chosen to reproduce the $m$th derivative at $\eta=0$. This evaluation scheme and the required quantum circuit are illustrated in Fig.~\ref{fig1}(a).

Further technical details and explicit examples of the multi-parameter GPSR construction are provided in Appendix~\ref{app:GPSR_response}. In the next section, we demonstrate the full workflow on the XXZ model by computing multi-time response functions via the GPSR and converting them to frequency-domain susceptibilities using standard Fourier processing.

\vspace{0.2cm}
\noindent\textbf{Hardware demonstration of quantum spin-chain response functions}

\noindent In this section, we use a single XXZ spin-chain model to illustrate our approach: Fig.~\ref{fig2} presents a noiseless simulation that benchmarks the extraction of higher-order response signals, and Fig.~\ref{fig3} then demonstrates the same reconstruction on a superconducting quantum processor. This organization makes the role of the hardware data transparent by providing a clean theoretical reference for comparison. We briefly outline the calculation of the higher-order response function for a quantum spin chain with a pulsed driving field (single- or two-pulse protocols in our experiments) \cite{huang2022variational,kokcu2024linear}. Consider the $N$-site spin$-\frac{1}{2}$ $XXZ$ chain \cite{sachdev1999quantum}, with Hamiltonian 
$
\mathbf H_{\text{XXZ}} = \frac{1}{4}\sum_{\langle i, j \rangle} \left( X_i X_j + Y_i Y_j + \Delta Z_iZ_j \right) - \frac{h_e}{2} \sum_i Z_i$, 
where $\Delta$ controls the anisotropy between the in-plane ($x, y$) and out-of-plane ($z$) spin interactions, and $h_e$ is the external magnetic field strength. 

We study both single-site and two-site (bond) observables $\mathbf A$, and the pump generator is chosen in the form $\mathbf B=\sum_{i=0}^{N-1} f_i X_i$. Different choices of $f_i$ correspond to different physical drivings; for instance, a momentum-selective instantaneous drive can be realized by setting $f_i=\cos(ki)$ with $k=\frac{2\pi m}{N}$, $m\in\mathbb{Z}$ \cite{kokcu2024linear}.

We begin with the single-channel, single-pulse setting (used in Fig.~\ref{fig2} and Fig.~\ref{fig3}(a)--(d)), where the pulse amplitude is a single parameter $\eta$.
In this case, the channel multi-index $(a_1,\dots,a_m)$ in $\chi^{(m)}_{\mathbf A;a_1\cdots a_m}$  takes the same value. As shown in Fig.~\ref{fig2}, we carry out a noiseless simulation that benchmarks the GPSR-based reconstruction of higher-order response contributions from the circuit-measured signals. Concretely, we evaluate $\langle \mathbf A(t)\rangle_{\eta}$ at the GPSR shift points and reconstruct $\{A^n(t)\}_{n=0}^{7}$ using the GPSR estimator. We additionally report auxiliary physical diagnostics (e.g., entanglement measures) across the pump strength $\eta$ and the anisotropy $\Delta$; while not required for the reconstruction itself, they help interpret the behavior near the phase boundary \cite{laflorencie2016quantum}. Here we take $N=12$ and find that results remain stable under moderate variations in the time grid and Trotter steps.

Building on the above noiseless reference, we next demonstrate the reconstruction on a real device. We demonstrate our algorithm on the 156-qubit superconducting processor \verb|ibm_kobe|, using a 12-qubit sub-chain that matches the nearest-neighbour XXZ topology and is chosen to minimize cumulative readout and two-qubit gate errors. Circuits are generated in the qclab framework and compiled with the F3C optimizer~\cite{camps2022algebraic,keip2025qclab,kokcu2022algebraic}. Time evolution of the $N=12$ XXZ chain (with $\Delta=0$ and longitudinal field $h_e=0.75$) is implemented via a 10-step first-order Trotterization of the Hamiltonian. Experiments sample $t\in[0,5]$ on a uniform grid of $51$ time points and use the discrete set of pump amplitudes $
\eta \in \{-\pi/4,\,0,\,\pi/4\}$,
required by the GPSR construction, with $8192$ shots per $(t,\vec \eta)$ configuration (with $\vec\eta=\eta$ for single-pulse and $\vec\eta=(\eta_0,\eta_1)$ for two-pulse). To facilitate a direct comparison, we compute the corresponding noiseless results using the same compiled $10$-step first-order Trotter circuits and display them as reference curves in Fig.~\ref{fig3}(a)–(f), so that the remaining deviation highlights hardware noise and finite-sampling effects. For the hardware runs we adopt a local pump choice (e.g., $B=X_3$), which simplifies the implementation while keeping the physical content of the response extraction unchanged.

In the spin-$\tfrac{1}{2}$ XXZ chain we adopt standard Pauli-operator conventions. The local longitudinal magnetization at site $j$ is defined as $M_z(j,t) := \langle Z_j(t)\rangle$, and the magnetization on a two-site region $\{j,j+1\}$ is
$
M_z^{(2)}(j,j+1;t) := \big\langle Z_j(t)+Z_{j+1}(t)\big\rangle$.
We also consider the local spin-current observable on the bond $(j,j+1)$,
$
J(j,j+1;t) := \big\langle X_j(t)Y_{j+1}(t) - Y_j(t)X_{j+1}(t)\big\rangle$,
which corresponds to the standard XXZ current operator up to an overall prefactor.

In Fig.~\ref{fig3}(a)–(f), we evaluate these observables near sites $(3,4)$, using $M_z^{(2)}(3,4;t)$ (two-site magnetization), $J(3,4;t)$ (spin current), and $\langle X_3(t)\rangle$ (transverse spin). Fig.~\ref{fig3}(a)–(d) present the fourth-order response in the time (a,c) and frequency (b,d) domains under a single-pulse protocol (pump $\mathbf B=X_3$ at $t=0$) for $\mathbf A=Z_3+Z_4$ and $\mathbf A=X_3Y_4-Y_3X_4$, while Fig.~\ref{fig3}(e),(f) show the fifth-order two-pulse response in the time (e) and frequency (f) domains for $\mathbf A=X_3$ (identical pumps at $t=0,1$). The two-pulse case can be viewed as a straightforward extension of the single-pulse setting: the reconstruction uses the same GPSR idea with $\vec\eta=(\eta_0,\eta_1)$, now accessing a higher-order signal through two pump times. An empirical exponential envelope $a(t)=\alpha e^{-\gamma t}$ is used to post-process the measured noisy spectra. We see that the nonlinear spectra after post-processing are consistent with the noiseless reference.

\vspace{0.2cm}
\noindent\textbf{High-order response in QSLs with 2DCS}
~\\
Having established and validated our GPSR-based framework on the XXZ chain, we now turn to quantum spin liquids (QSLs) and their signatures in two-dimensional coherent spectroscopy (2DCS). As is well known, quasi-particle excitation spectra can be inferred from nonlinear spectroscopic measurements. In recent years, the emergence of pump--probe experiments \cite{hamm2005principles,norman2018principles} and two-dimensional coherent spectroscopy (2DCS) techniques \cite{woerner2013ultrafast,lu2017coherent,mahmood2021observation,wan2019resolving,choi2020theory,parameswaran2020asymptotically,nandkishore2021spectroscopic} has enabled the exploration of complex materials; these experimental methods are also well-established tools in cold-atom physics and chemistry \cite{cai2020quantum,huang2022variational}. However, most theoretical studies have focused on linear response quantities, and the spectroscopy of QSLs remains far more developed within linear-response frameworks than in the genuinely nonlinear regime. While recent nonlinear studies have begun to uncover signatures of fractionalized excitations and their interactions, such efforts remain relatively recent and largely confined to specific models and protocols \cite{qi2009dynamics,punk2014topological,knolle2014raman,knolle2014dynamics,knolle2015dynamics,kamfor2014spectroscopy,morampudi2020spectroscopy,wu2024selective,pfender2019high}. Here we develop a quantum algorithm to access high-order nonlinear responses in strongly correlated many-body systems.

We focus on quantum spin liquids (QSLs), whose fractionalized excitations can manifest in nonlinear spectroscopic signals. As a concrete example, we investigate the 2D Kitaev toric code \cite{kitaev2003fault,kanega2021linear} and analyze its pump--probe response, where the system is perturbed by two pulses with a tunable time delay between them. In this setting, anyonic excitations created by the initial pulse can propagate and interact nontrivially with those generated by the subsequent pulse, potentially undergoing braiding processes that imprint characteristic phase factors in nonlinear signals.

We analyze the Kitaev toric code using the same GPSR-based simulation and analysis pipeline as introduced for the XXZ chain:
$
	\mathbf H_{\mathrm{TC}} = -J_A \sum_{v} \prod_{j\in v} X_j \;-\; J_B \sum_{p} \prod_{j\in p} Z_j$.
In the toric code, qubits reside on lattice edges $j$; $X_j, Z_j$ act on those edges, with the star and plaquette operators $\mathbf A_v \equiv \prod_{j \in v} X_j$ and $\mathbf B_p \equiv \prod_{j \in p} Z_j$ around vertex $v$ and plaquette $p$, respectively. 

\begin{figure*}[!htbp]
	\centering
	\includegraphics[width=0.95\textwidth]{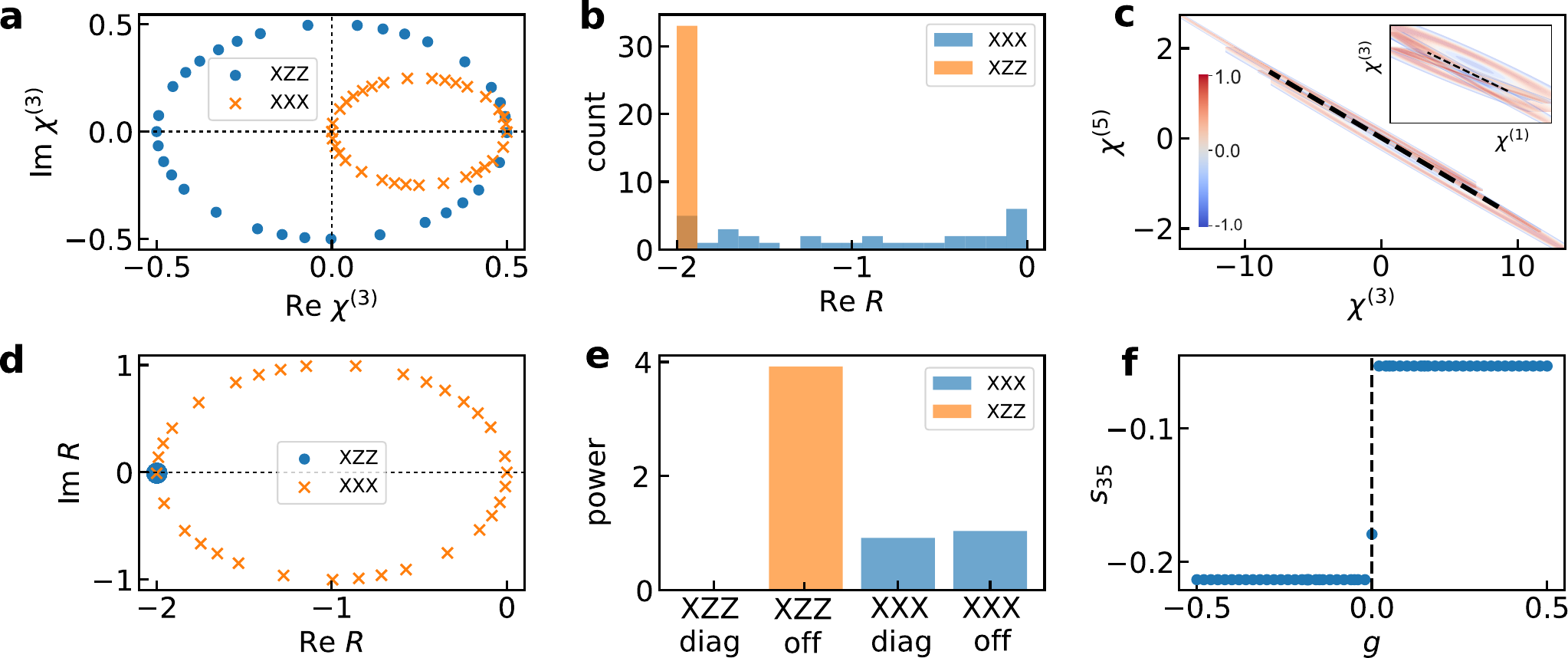}
\caption{Quasi-particle signatures from nonlinear pump--probe contributions in the Kitaev toric-code model. (a) Complex-plane trajectories of the reconstructed third-order contribution $C^{(3)}$ for the two channels (XZZ and XXX), showing distinct phase winding in the complex plane. (b) Histogram of the real part of the complex channel-contrast ratio $R$ (defined in the main text), demonstrating a clear separation between the two channels. (c) Kernel-density estimate (KDE) of the joint distribution of $(C^{(3)},C^{(5)})$ at a representative coupling $g$; the dashed line is the principal-axis (PCA) fit whose slope defines $s_{35}$ (with an inset showing the analogous $(C^{(1)},C^{(3)})$ correlation). (d) Complex-plane points of $R$, further visualizing the channel-dependent phase structure. (e) Frequency-domain “diagonal vs off-diagonal” spectral weight (integrated power) obtained from the 2D Fourier spectrum at $\Delta\Omega=0$, contrasting diagonal ($\omega_1\approx\omega_2$) and off-diagonal contributions for XZZ and XXX. (f) Extracted $s_{35}(g)$ as a function of coupling $g$; the sharp change across $g=0$ (dashed vertical line) indicates the transition point in the coupling sweep.}
	\label{fig4}
\end{figure*}

We study the pump--probe protocol as follows. An impulsive pump at $t=0$ applies a unitary
rotation $e^{-i\eta\,\mathbf B}$
to the toric-code ground state $|\psi_0\rangle$, where $\mathbf B$ is a chosen Pauli-string operator
that specifies the pump channel (e.g., $\mathrm{XZZ}$ or $\mathrm{XXX}$). For the same channel, we choose two probe Pauli-string operators $\mathbf A_1$ and $\mathbf A_2$
specified by two edge strings in our implementation (denoted by $\gamma_1$ and $\gamma_2$), and define the
composite probe observable
$
\mathbf A(t_1,t_2)=\mathbf A_2(t_1{+}t_2)\,\mathbf A_1(t_1)
$, where $\mathbf A_1(t)$ and $\mathbf A_2(t)$ are interaction-picture operators with respect to the unperturbed toric-code Hamiltonian $\mathbf H_{\mathrm{TC}}$.
For each time pair $(t_1,t_2)$ we evaluate the pump-dependent two-time correlator and expand it in the pump amplitude $\eta$ as
$
C(t_1,t_2;\eta)
=
\langle \psi_0 |\,e^{i\eta\mathbf B}\,
\mathbf A(t_1,t_2) \,
e^{-i\eta\mathbf B}\,| \psi_0 \rangle =\sum_{n\ge 0} C^{(n)}(t_1,t_2,\eta)$, where $C^{(n)}$ denotes the $n$th-order contribution in the pump-amplitude expansion.

Fig.~\ref{fig4}(a) shows that the complex-plane samples of $C^{(3)}(t_1,t_2)$ form clearly distinct
point clouds for the {XZZ} and {XXX} geometries, indicating qualitatively different third-order
structures. To directly expose the pump-induced phase signature, we construct the complex contrast ratio
$R(t_1,t_2)=C(t_1,t_2;\kappa)/C(t_1,t_2;0)-1$ at a fixed pump angle $\kappa=\pi/2$.
As shown in Fig.~\ref{fig4}(b,d), the {XZZ} channel concentrates near the expected offset
$-2=e^{i\pi}-1$, while the {XXX} reference remains distributed near $0$, providing distribution-level
evidence for a nontrivial phase contribution.

We quantify the frequency-domain signal by summing the spectral power near the diagonal
$\omega_1=\omega_2$ (i.e., $|\omega_1-\omega_2|\approx 0$) and comparing it with the off-diagonal
weight. The integrated diagonal/off-diagonal powers in Fig.~\ref{fig4}(e) cleanly distinguish the two geometries and provide a compact spectral diagnostic of cross-peak-like weight.

Finally, we probe coupling-driven changes in the toric-code pump--probe response by fixing $J_A$
and sweeping $g\equiv J_B$ through $g=0$. For each $g$, we sample $(t_1,t_2)$ at a fixed pump amplitude
and treat $(C^{(1)},C^{(3)})$ and $(C^{(3)},C^{(5)})$ as point clouds. Their cross-order
correlations are quantified by the first principal-axis (total least-squares) slopes $s_{13}(g)$ and
$s_{35}(g)$. Fig.~\ref{fig4}(c) shows a representative Kernel-density estimate (KDE) of $(C^{(3)},C^{(5)})$ with its
principal axis (inset: $(C^{(1)},C^{(3)})$), demonstrating an approximately affine trend.
Notably, $s_{35}(g)$ in Fig.~\ref{fig4}(f) exhibits a sharp step-like change as $g$ crosses $0$,
yielding a clean transition signature. The higher-order slope provides markedly stronger contrast
than lower-order correlations, indicating that high-order responses more sensitively encode the
underlying excitation/interference (quasiparticle/anyon) structure.

\begin{table*}
\centering
\caption{\textbf{Summary of computed examples in the appendices.} This table compiles the observables evaluated in this work and their locations in the Appendices (sections/figures/equations). Entries cover the multi-time higher-order response and GPSR construction, direct time-/frequency-domain extraction schemes, magnetization response, two- and four-point spin correlations, spin current density, and quasi-particle excitation spectra.}
\label{tab:appendix-summary}
\begin{tabular}{|l|l|l|}
\hline
\textbf{Example / Scenario} & \textbf{Observable / Response} & \textbf{Reference (App./Fig./Eq.)} \\
\hline
Multi-time higher-order response & $\chi^{(m)}_{\mathbf A;a_1\cdots a_m}(t;t_1,\dots,t_m)$;\ generating functional $\mathcal G_{\mathbf A}(\vec\eta;t)$ & App.~A;\ Eqs.~\eqref{eq:derivative-to-commutator}, \eqref{eq:chi-psr-unified} \\
\hline
GPSR & Shift set $\{s_{a,p}\}$ and weights $c_{a,p}^{(r)}$ & App.~B;\ Eqs.~\eqref{eq:psr-single-system}--\eqref{eq:chi-psr-unified} \\
\hline
Direct construction/extraction of $\chi^{(n)}$ & Time-/frequency-domain schemes & App.~C;\ Figs.~\ref{a}--\ref{b} \\
\hline
Magnetization response & $\langle \mathbf M^x(t)\rangle_{\eta},\ \langle \mathbf M^y(t)\rangle_{\eta}$ & App.~D;\ Figs.~\ref{6}--\ref{7} \\
\hline
Spin correlations & $C^{\alpha\beta}_{i,j}(t)$;\ $C^{\alpha\beta\gamma\delta}_{i,j,k,l}(t)$ (e.g.\ $C^{xyxz}_{0,1,2,3}(t)$) & App.~V;\ Figs.~\ref{8}--\ref{10} \\
\hline
Spin current density & $J^{a}_{s,ij}(t)$\ $(a=x,y,z)$ & App.~F;\ Figs.~\ref{11}--\ref{13} \\
\hline
Quasi-particle excitation & Spectra / linear components (e.g.\ PCA slope $m_{\mathrm{PCA}}$) & App.~G;\ Figs.~\ref{14}--\ref{16} \\
\hline
\end{tabular}
\end{table*}

\begin{figure}[!htbp]
	\centering
	\includegraphics[width=0.48\textwidth]{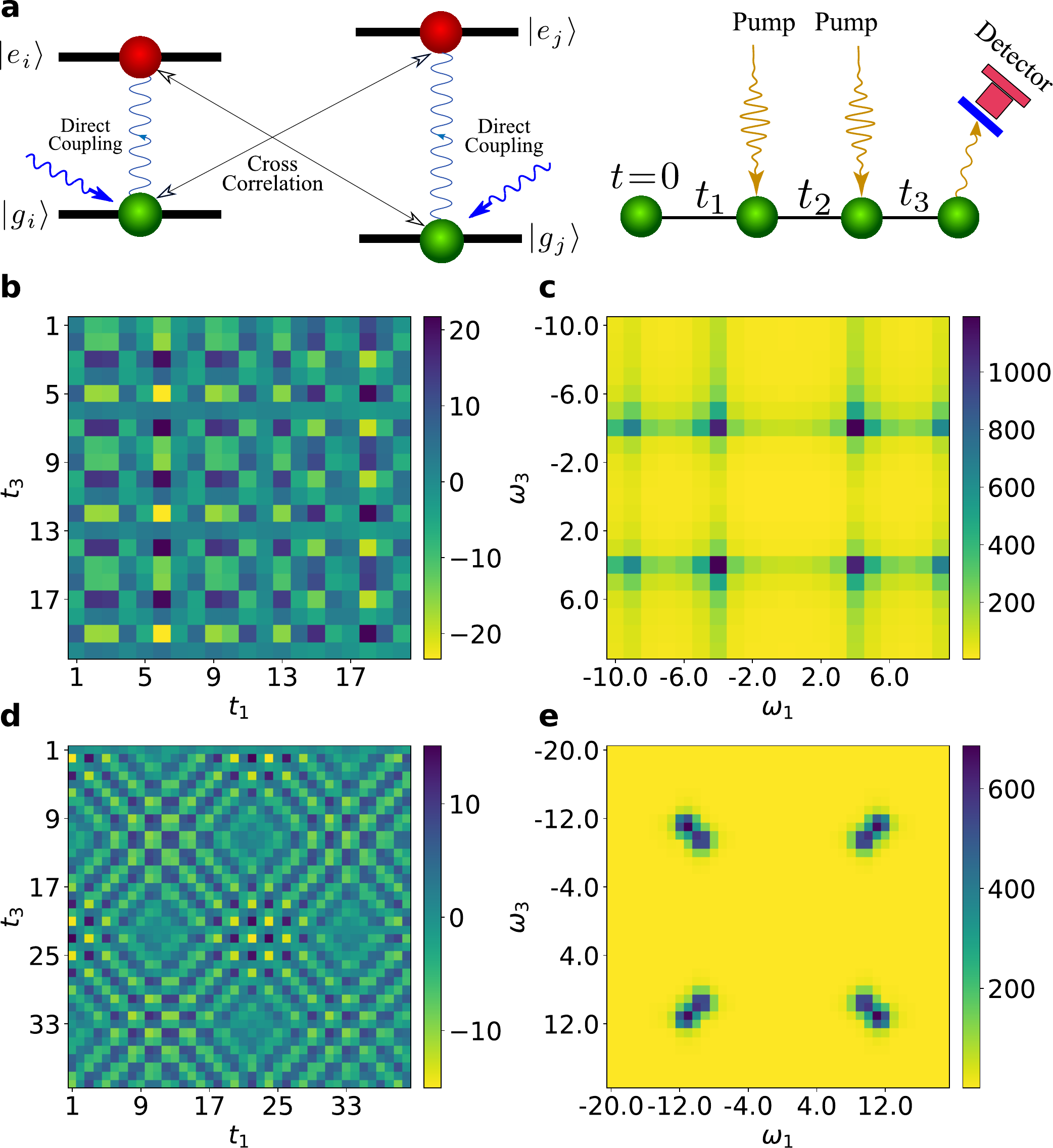}
\caption{Cross-peak signatures in coupled two-level systems (TLSs) probed by two-dimensional optical spectroscopy (2DOS). (a) Schematic of the three-pulse protocol for probing the third-order response. (b),(c) Direct-coupling case: time-domain signal $S^{(3)}(t_1,t_3)$ and the corresponding two-dimensional spectrum $S^{(3)}(\omega_1,\omega_3)$ for $\mathbf H_{\mathrm{TLSs}}$ with $(\omega_1,\omega_2,J)=(0.5,1.0,0.8)$. (d),(e) Indirect-coupling case: $S^{(3)}(t_1,t_3)$ and the corresponding spectrum for the shared-mode Hamiltonian $\mathbf H_{\mathrm{SB}}$ with $(\omega_1,\omega_2,\omega_b,g)\approx(2.64,\,3.20,\,0.04,\,0.05)$. In both cases, off-diagonal cross-peaks indicate correlations between distinct transitions.}
	\label{fig5}
\end{figure}

\vspace{0.2cm}
\noindent\textbf{Cross-peaks in TLS with 2DOS} 
~\\
Two-dimensional optical spectroscopy (2DOS) is a third-order nonlinear spectroscopic technique with high spectral resolution that probes absorption dynamics \cite{cheung2025quantum}. In a standard protocol, three ultrafast coherent pulses interact sequentially with the sample, and the detected third-order signal is double–Fourier transformed to yield the 2DOS. In the resulting $(\omega_1,\omega_3)$ spectrum, diagonal features are associated with single-transition resonances, whereas off-diagonal cross-peaks signal correlations between distinct transitions. Here we demonstrate cross-peaks formation in two minimal TLS settings: (i) a coherently exchange-coupled dimer with direct inter-TLS interaction, and (ii) a minimal indirect-coupling model where the two TLSs develop correlations via a shared mode, despite having no direct energy-exchange term.

We first consider two coherently coupled TLSs with Hamiltonian
$
\mathbf H_{\text{TLSs}}
=\sum_{i=0}^1 \frac{\omega_i}{2}Z_i
+J\big(X_0 X_1
+Y_0 Y_1
+Z_0 Z_1\big)$,
where $\omega_i$ are their transition energies, and $J$ is the coherent inter-TLS coupling encoded by the $XX$, $YY$, and $ZZ$ interaction terms. This direct coupling hybridizes the bare TLS transitions and yields prominent cross-peaks in 2DOS.

To contrast with the direct-coupling case, we also simulate an indirect-coupling setting in which the two TLSs have no direct interaction, but are correlated through a shared mode. A minimal Hamiltonian reads
$
\mathbf H_{\text{SB}}
=\sum_{i=0}^1 \frac{\omega_i}{2}Z_i
+\frac{\omega_b}{2}Z_b
+g\big(Z_0 X_b+Z_1 X_b\big)$,
where $\omega_b$ is the mode frequency and $g$ is the coupling strength. This model generates cross-peaks via correlation-induced pathways rather than population transfer between the TLSs.

With the interaction-picture operators $\mathbf A(t)=e^{i\mathbf H_0 t}\mathbf A e^{-i\mathbf H_0 t}$ and $\mathbf B(t)=e^{i\mathbf H_0 t}\mathbf B e^{-i\mathbf H_0 t}$, and fixing the waiting time $t_2$, the third-order response function for the ordered pulse sequence is $\chi^{(3)}(t_1,t_3)=i^3\,\langle\psi_0|[[[\mathbf A(t_1+t_2+t_3), \mathbf B(t_1+t_2)],\mathbf B(t_1)], \mathbf B(0)]|\psi_0\rangle$. For visualization, we plot its absorptive component $S^{(3)}(t_1,t_3)\equiv \operatorname{Im}\,\chi^{(3)}(t_1,t_3)$.
In our simulations, the detection operator is chosen as the total polarization $\mathbf A=X_0+X_1$, and the pump generator is
$\mathbf B=\sum_{i=0}^{1}\cos(ks\,i)\,X_i$ with $N=2$, where $ks$ controls the relative phase of the two pumped channels (in particular, $ks=0$ corresponds to in-phase pumping).
To identify cross-peaks, we compute the two-dimensional spectrum $S^{(3)}(\omega_1,\omega_3)$ as the discrete double Fourier transform of $S^{(3)}(t_1,t_3)$ over $t_1$ and $t_3$. The resulting 2DOS exhibits clear off-diagonal cross-peaks in both settings: in the direct-coupling model, their spectral weights and locations are governed by the hybridized transitions controlled by $J$, whereas in the indirect-coupling model, cross-peaks arise from correlation-induced pathways mediated by the shared mode even in the absence of a direct inter-TLS exchange term.\\

\noindent\textbf{Discussion}

\noindent 
Our work demonstrates that nonlinear response functions can be accessed on quantum simulators through the finite reconstruction of experimentally measurable driven signals. By expressing higher-order response coefficients as mixed derivatives with respect to pump amplitudes and reconstructing them using the GPSR, nonlinear spectroscopic signals can be obtained from a finite set of expectation-value measurements at shifted control parameters. In this way, the framework replaces the explicit evaluation of multi-time operator expansions by a measurement-based reconstruction procedure that fits naturally into standard quantum-simulation workflows.

This perspective has practical implications for quantum hardware implementations. Because the protocol relies only on real-time evolution and repeated expectation-value measurements, it avoids the ancillary registers and controlled multi-time operations that often appear in circuit constructions of higher-order correlation functions. As a result, nonlinear responses can be extracted using circuits that remain close to conventional dynamical-simulation circuits, making the approach suitable for current and near-term quantum simulation platforms.

The examples studied in this work further illustrate how nonlinear-response measurements can reveal dynamical information beyond that captured by linear probes. In interacting spin systems, higher-order signals encode correlations associated with collective excitations and interaction-driven dynamics. In topologically ordered models, pump--probe protocols can probe interference processes of emergent quasiparticles. Similarly, in minimal optical models, the reconstructed multidimensional spectra reproduce cross-peak structures that reflect correlations between distinct transitions. Taken together, these examples show that the framework is not limited to a specific observable or model class, but instead provides a common route for extracting nonlinear dynamical signatures across different physical platforms.

The measurement cost of the method is determined primarily by the spectral structure of the pump generators, which sets the number of parameter shifts required in the GPSR reconstruction. For many physically relevant operators—including local Pauli terms, few-body observables, or collective polarization operators—the number of distinct spectral gaps remains small, implying that the number of circuit evaluations does not show intrinsic exponential growth with system size at fixed response order. Under these conditions, the practical overhead remains manageable for physically relevant nonlinear-response protocols, without the heavy circuit cost associated with explicit multi-time constructions.

Taken together, these results suggest that programmable quantum simulators may serve not only as tools for computing equilibrium properties but also as platforms for implementing nonlinear spectroscopic probes of quantum matter. More broadly, the present framework provides a hardware-compatible route for studying nonlinear dynamical properties of quantum matter beyond the linear-response regime. Extending this framework to more complex pulse sequences, multidimensional spectroscopy protocols, or nonlinear transport measurements represents a promising direction for future investigations. Further nonlinear examples are provided in the Appendix, with a summary in Table~\ref{tab:appendix-summary}.\\

\noindent\textbf{Acknowledgement}\\
\noindent This work is supported by 
Quantum Science and Technology-National Science and Technology Major Project (2023ZD0300200), 
the National Natural Science Foundation of China Grant (No.~12361161602, 22303005), 
NSAF (Grant No.~U2330201), 
Beijing Natural Science Foundation Z250004, 
Beijing Science and Technology Planning Project (Grant No.~Z25110100810000), the
Startup Fund of the Institute of High Energy Physics
and the High-performance Computing Platform of Peking University. 
X.W. is supported by the RIKEN TRIP initiative (RIKEN Quantum) and the UTokyo Quantum Initiative.

\bibliography{main.bib}


\clearpage          
\onecolumngrid      

\section*{Appendix}
\section*{Appendix A: Multi-time Higher-order Response and Generalized Parameter-Shift Rule}
\label{app:GPSR_response}

This appendix derives a unified bridge from multi-time, higher-order response theory to the generalized parameter-shift rule (GPSR) used in our circuit implementations.
We start from the interaction-picture Kubo expansion: the $m$th-order response kernel is a nested-commutator object with explicit causality.
We then recast the response as derivatives of a generating functional with respect to the driving fields.
Restricting the drive to a finite set of $\delta$-pulses reduces the functional dependence on continuous fields to an ordinary dependence on a finite parameter vector $\vec\eta$ that is directly programmable on hardware.
When each pulse generator has a finite discrete spectrum, the generating function becomes a finite Fourier series in the corresponding control parameter(s), which enables an exact GPSR: any (mixed) derivative at $\vec\eta=\vec 0$ can be reconstructed from finitely many shifted evaluations at discrete parameter configurations.
This provides an experimentally accessible route from formal higher-order response theory to finite-shot estimation on quantum devices.

\subsection*{A.1 General form of the response function}

We consider an unperturbed Hamiltonian $\mathbf H_0$ and an observable $\mathbf A$.
In the interaction picture,
\begin{align}
	\mathbf A(t)=U_0^\dagger(t)\,\mathbf A\,U_0(t),\qquad U_0(t)=e^{-i\mathbf H_0 t}.
\end{align}
The system is driven by a weak, linearly coupled perturbation
\begin{align}
	\mathbf H_I(t)=\sum_{a} h_a(t)\,\mathbf B_a(t),\qquad \mathbf B_a(t)=U_0^\dagger(t)\,\mathbf B_a\,U_0(t),
\end{align}
where $a$ labels a finite set of drive ``channels'' (e.g., different polarization/spin components or different Pauli strings in circuits).
The initial state is a (possibly mixed) density matrix
\begin{align}
	\boldsymbol{\rho}_0,\qquad \langle X\rangle_0:=\mathrm{Tr}(X\boldsymbol{\rho}_0).
\end{align}
We use the commutator superoperator
\begin{align}
	\operatorname{ad}_X(Y)=[X,Y].
\end{align}

In the interaction picture, the density matrix satisfies
\begin{align}
	i\,\frac{d}{dt}\boldsymbol{\rho}_I(t)=[\mathbf H_I(t),\boldsymbol{\rho}_I(t)],\qquad \boldsymbol{\rho}_I(t_0)=\boldsymbol{\rho}_0.
\end{align}
Taking $t_0\to-\infty$, the integral equation reads
\begin{align}
	\boldsymbol{\rho}_I(t)=\boldsymbol{\rho}_0-i\int_{-\infty}^{t}dt_1\,[\mathbf H_I(t_1),\boldsymbol{\rho}_I(t_1)].
	\label{eq:rhoI-integral}
\end{align}
Iterating Eq.~\eqref{eq:rhoI-integral} yields the Dyson series.
In general, the $m$th-order contribution is
\begin{align}
	\boldsymbol{\rho}_I^{(m)}(t)
	= (-i)^m \int_{-\infty}^{t}dt_1\int_{-\infty}^{t_1}dt_2\cdots\int_{-\infty}^{t_{m-1}}dt_m\,
	\operatorname{ad}_{\mathbf H_I(t_1)}\cdots\operatorname{ad}_{\mathbf H_I(t_m)}\boldsymbol{\rho}_0.
	\label{eq:rhoI-mth}
\end{align}
Using step functions, the nested limits in Eq.~\eqref{eq:rhoI-mth} can be recast as
\begin{align}
	\int_{-\infty}^{t}dt_1\int_{-\infty}^{t_1}dt_2\cdots\int_{-\infty}^{t_{m-1}}dt_m
	=\int dt_1\cdots dt_m\,\theta(t-t_1)\theta(t_1-t_2)\cdots\theta(t_{m-1}-t_m).
\end{align}
Thus
\begin{align}
	\boldsymbol{\rho}_I(t)-\boldsymbol{\rho}_0
	= \sum_{m=1}^{\infty}(-i)^m\int dt_1\cdots dt_m\,
	\theta(t-t_1)\cdots\theta(t_{m-1}-t_m)\,
	\operatorname{ad}_{\mathbf H_I(t_1)}\cdots\operatorname{ad}_{\mathbf H_I(t_m)}\boldsymbol{\rho}_0.
	\label{eq:rhoI-dyson}
\end{align}

The induced deviation of $\langle \mathbf A(t)\rangle$ is
\begin{align}
	\delta\langle \mathbf A(t)\rangle
	=\mathrm{Tr}\{\mathbf A(t)[\boldsymbol{\rho}_I(t)-\boldsymbol{\rho}_0]\}.
	\label{eq:deltaA-def}
\end{align}
Substituting Eq.~\eqref{eq:rhoI-dyson} into Eq.~\eqref{eq:deltaA-def} gives
\begin{align}
	\delta\langle \mathbf A(t)\rangle
	=\sum_{m=1}^{\infty}(-i)^m\int dt_1\cdots dt_m\,
	\theta(t-t_1)\cdots\theta(t_{m-1}-t_m)\,
	\mathrm{Tr}\{\mathbf A(t)\,\operatorname{ad}_{\mathbf H_I(t_1)}\cdots\operatorname{ad}_{\mathbf H_I(t_m)}\boldsymbol{\rho}_0\}.
	\label{eq:deltaA-series}
\end{align}
We now use the trace identity
\begin{align}
	\mathrm{Tr}\{X[Y,\boldsymbol{\rho}_0]\}=\mathrm{Tr}\{[X,Y]\boldsymbol{\rho}_0\},
	\label{eq:trace-identity}
\end{align}
which moves commutators from $\boldsymbol{\rho}_0$ onto $\mathbf A(t)$.
By induction,
\begin{align}
	\mathrm{Tr}\{\mathbf A\,\operatorname{ad}_{\mathbf H_I(t_1)}\cdots\operatorname{ad}_{\mathbf H_I(t_m)}\boldsymbol{\rho}_0\}
	=(-1)^m\,\mathrm{Tr}\{\operatorname{ad}_{\mathbf H_I(t_m)}\cdots\operatorname{ad}_{\mathbf H_I(t_1)}\mathbf A\,\boldsymbol{\rho}_0\}.
	\label{eq:trace-ad-identity}
\end{align}
Inserting Eq.~\eqref{eq:trace-ad-identity} into Eq.~\eqref{eq:deltaA-series} yields the causal (time-ordered) Kubo form
\begin{align}
	\delta\langle \mathbf A(t)\rangle
	=\sum_{m=1}^{\infty} i^m\int dt_1\cdots dt_m\,
	\theta(t-t_1)\cdots\theta(t_{m-1}-t_m)\,
	\big\langle \operatorname{ad}_{\mathbf H_I(t_m)}\cdots\operatorname{ad}_{\mathbf H_I(t_1)}\mathbf A(t)\big\rangle_0.
	\label{eq:delta-A(x)}
\end{align}

Expanding $\mathbf H_I(t_k)=\sum_{a_k} h_{a_k}(t_k)\mathbf B_{a_k}(t_k)$ in Eq.~\eqref{eq:delta-A(x)}, we obtain the standard response expansion
\begin{align}
	\delta\langle \mathbf A(t)\rangle
	=\sum_{m=1}^{\infty}\sum_{a_1,\dots,a_m}
	\int dt_1\cdots dt_m\;
	\chi^{(m)}_{\mathbf A;a_1\cdots a_m}(t;t_1,\dots,t_m)\,
	\prod_{k=1}^{m} h_{a_k}(t_k),
	\label{eq:delta-A-symmetrized}
\end{align}
with the causal $m$th-order kernel
\begin{align}
	\boxed{
		\chi^{(m)}_{\mathbf A;a_1\cdots a_m}(t;t_1,\dots,t_m)
		=
		i^m\,
		\theta(t-t_1)\cdots\theta(t_{m-1}-t_m)\,
		\big\langle
		\operatorname{ad}_{\mathbf B_{a_m}(t_m)}\cdots\operatorname{ad}_{\mathbf B_{a_1}(t_1)}\mathbf A(t)
		\big\rangle_0.
	}
	\label{eq:chi-kubo-causal}
\end{align}

The Fourier transform (convention $f(\omega)=\int dt\,e^{i\omega t}f(t)$) is
\begin{align}
	\chi^{(m)}_{\mathbf A;a_1\cdots a_m}(\omega_1,\dots,\omega_m)
	=\int dt_1\cdots dt_m\;
	e^{i\sum_k\omega_k t_k}\,
	\chi^{(m)}_{\mathbf A;a_1\cdots a_m}(t;t_1,\dots,t_m),
\end{align}
which enters nonlinear spectra such as multidimensional spectroscopy.

For a given field configuration $\vec h=\{h_a(t)\}$, define the generating functional
\begin{align}
	\mathcal G_{\mathbf A}[\vec h;t]\equiv \langle \mathbf A(t)\rangle_{\vec h}.
\end{align}
Formally expanding around $\vec h=\vec 0$ gives
\begin{align}
	\mathcal G_{\mathbf A}[\vec h;t]-\langle \mathbf A(t)\rangle_0
	=\sum_{m=1}^\infty\frac{1}{m!}\sum_{a_1,\dots,a_m}\int dt_1\cdots dt_m\,
	\Gamma^{(m)}_{\mathbf A;a_1\cdots a_m}(t;t_1,\dots,t_m)\,
	\prod_{k=1}^m h_{a_k}(t_k),
\end{align}
where
\begin{align}
	\Gamma^{(m)}_{\mathbf A;a_1\cdots a_m}(t;t_1,\dots,t_m)
	=
	\left.
	\frac{\delta^m\langle \mathbf A(t)\rangle_{\vec h}}
	{\delta h_{a_1}(t_1)\cdots\delta h_{a_m}(t_m)}
	\right|_{\vec h=\vec 0}.
	\label{eq:Gamma-functional-derivative}
\end{align}
The $m!$ and the precise symmetry of $\Gamma^{(m)}$ depend on whether one treats $(t_1,\dots,t_m)$ as unordered arguments (fully symmetric functional derivatives) or restricts to an explicit time-ordered sector (where no combinatorial factor appears).
Throughout this appendix, we keep the causal Kubo kernel $\chi^{(m)}$ in Eq.~\eqref{eq:chi-kubo-causal} as the primary object, and use Eq.~\eqref{eq:Gamma-functional-derivative} only as a device to connect response coefficients to derivatives of a finite-dimensional parameter function under $\delta$-pulse drives (Sec.~3).

\subsection*{A.2 $\delta$-pulse restriction and finite-dimensional parameterization}

We now restrict the drive to a finite set of $\delta$-pulses, as implemented in our circuits.
Throughout the paper we group pulses into $L$ drive channels labeled by $a\in\{0,1,\dots,L-1\}$.
Channel $a$ has a single programmable amplitude $\eta_a$ but may contain multiple pulses at times $\{t_{ab}\}_{b=0}^{n_a-1}$:
\begin{align}
	\mathbf H_I(t)
	=\sum_{a=0}^{L-1} h_a(t)\,\mathbf B_a(t),\qquad
	h_a(t)=\eta_a\sum_{b=0}^{n_a-1}\delta(t-t_{ab}).
	\label{eq:channel-pulse-Hamiltonian}
\end{align}
Under this $\delta$-pulse restriction, the generating functional $\mathcal G_{\mathbf A}[\vec h;t]$ reduces to an ordinary function of the finite parameter vector
$\vec\eta=(\eta_0,\dots,\eta_{L-1})$:
\begin{align}
	\mathcal G_{\mathbf A}(\vec\eta;t)
	:=\mathcal G_{\mathbf A}[\vec h^{(\vec\eta)};t]
	=\langle \mathbf A(t)\rangle_{\vec\eta}.
	\label{eq:G-finite-eta}
\end{align}

\paragraph{Response expansion under channel $\delta$-pulses.}
Substituting Eq.~\eqref{eq:channel-pulse-Hamiltonian} into the Volterra/Kubo expansion Eq.~\eqref{eq:delta-A-symmetrized} gives a polynomial in $\{\eta_a\}$:
\begin{align}
	\delta\langle \mathbf A(t)\rangle
	&=
	\sum_{m=1}^{\infty}
	\sum_{a_1,\dots,a_m=0}^{L-1}
	\Bigg[
	\sum_{b_1=0}^{n_{a_1}-1}\cdots\sum_{b_m=0}^{n_{a_m}-1}
	\chi^{(m)}_{\mathbf A; a_1\cdots a_m}
	\big(t; t_{a_1b_1},\dots,t_{a_mb_m}\big)
	\Bigg]
	\prod_{k=1}^m \eta_{a_k}.
	\label{eq:delta-A-channel-delta}
\end{align}

\paragraph{$\beta_a$-counting and the correct Taylor prefactor.}
For a given term indexed by $(a_1,\dots,a_m)$, define the multiplicities (occupation numbers)
\begin{align}
	\beta_a:=\sum_{k=1}^{m}\delta_{a_k,a},\qquad \sum_{a=0}^{L-1}\beta_a=m,
	\label{eq:beta-definition}
\end{align}
so that the monomial satisfies $\prod_{k=1}^m\eta_{a_k}=\prod_{a=0}^{L-1}\eta_a^{\beta_a}$.
The multivariate Taylor expansion of $\mathcal G_{\mathbf A}(\vec\eta;t)$ around $\vec\eta=\vec 0$ reads
\begin{align}
	\mathcal G_{\mathbf A}(\vec\eta;t)-\mathcal G_{\mathbf A}(\vec 0;t)
	=\sum_{m=1}^{\infty}
	\ \sum_{\{\beta_a\}:\,\sum_a\beta_a=m}
	\frac{1}{\prod_{a=0}^{L-1}\beta_a!}
	\left.
	\Bigg(\prod_{a=0}^{L-1}\frac{\partial^{\beta_a}}{\partial\eta_a^{\beta_a}}\Bigg)
	\mathcal G_{\mathbf A}(\vec\eta;t)
	\right|_{\vec\eta=\vec 0}
	\prod_{a=0}^{L-1}\eta_a^{\beta_a}.
	\label{eq:taylor-channel-beta}
\end{align}
Comparing Eq.~\eqref{eq:delta-A-channel-delta} with Eq.~\eqref{eq:taylor-channel-beta}, we obtain the key identification
\begin{align}
	\sum_{b_1=0}^{n_{a_1}-1}\cdots\sum_{b_m=0}^{n_{a_m}-1}
	\chi^{(m)}_{\mathbf A; a_1\cdots a_m}
	\big(t; t_{a_1b_1},\dots,t_{a_mb_m}\big)
	=
	\frac{1}{\prod_{a=0}^{L-1}\beta_a!}
	\left.
	\Bigg(\prod_{a=0}^{L-1}\frac{\partial^{\beta_a}}{\partial\eta_a^{\beta_a}}\Bigg)
	\mathcal G_{\mathbf A}(\vec\eta;t)
	\right|_{\vec\eta=\vec 0}.
	\label{eq:chi-to-derivatives-channel}
\end{align}
The only subtlety relative to the pulse-resolved convention is precisely this $\beta_a$-dependent prefactor.
If every pulse carried an \emph{independent} amplitude parameter (no identifications across channels), one would recover the familiar $1/m!$ prefactor.

\subsection*{A.3 Generalized parameter-shift representation of response functions}

\subsubsection*{Single-parameter GPSR}

We now connect the derivative form in Eq.~\eqref{eq:chi-to-derivatives-channel} to GPSR.
Consider a single-parameter expectation value
\begin{align}
	F(\eta)=\big\langle e^{i\eta \mathbf B}\,\mathbf A\,e^{-i\eta \mathbf B}\big\rangle_0,
	\label{eq:F-def}
\end{align}
where $\mathbf B$ is a bounded Hermitian operator with a finite discrete spectrum
\begin{align}
	\mathbf B=\sum_{s\in S}\lambda_s P_s,\qquad
	P_sP_{s'}=\delta_{s,s'}P_s,\quad\sum_s P_s=\mathbb I.
\end{align}
Inserting the spectral decomposition into Eq.~\eqref{eq:F-def} yields a finite Fourier series
\begin{align}
	F(\eta)
	&=\sum_{s,s'} e^{i\eta(\lambda_s-\lambda_{s'})}\,\mathrm{Tr}\big[\boldsymbol{\rho}_0\,P_s \mathbf A P_{s'}\big]
	=\sum_{\omega\in\Omega}\hat F(\omega)\,e^{i\omega\eta},
	\label{eq:Fourier_expansion}
\end{align}
with frequency set
\begin{align}
	\Omega:=\{\lambda_s-\lambda_{s'}\mid s,s'\in S\},
\end{align}
and coefficients
\begin{align}
	\hat F(\omega)
	:=\sum_{(s,s'):\lambda_s-\lambda_{s'}=\omega}
	\mathrm{Tr}\big[\boldsymbol{\rho}_0\,P_s \mathbf A P_{s'}\big].
\end{align}
Therefore
\begin{align}
	F^{(r)}(0)=\sum_{\omega\in\Omega}\hat F(\omega)(i\omega)^r.
	\label{eq:F-r-th-derivative}
\end{align}

On hardware we estimate $F(\eta)$ at selected shifts $\{s_p\}_{p=1}^{M}$.
We seek coefficients $\{c_{p}^{(r)}\}$ such that
\begin{align}
	F^{(r)}(0)=\sum_{p=1}^{M} c_{p}^{(r)}\,F(s_p).
	\label{eq:psr-single-derivative}
\end{align}
Substituting Eq.~\eqref{eq:Fourier_expansion} into Eq.~\eqref{eq:psr-single-derivative} gives the linear system
\begin{align}
	\sum_{p=1}^{M} c_{p}^{(r)} e^{i\omega s_p}=(i\omega)^r,\qquad\forall\omega\in\Omega,
	\label{eq:psr-single-system}
\end{align}
whose solution defines the GPSR rule.
A minimal choice is typically $M\ge |\Omega|$ (with equality if the resulting Vandermonde-type matrix is invertible).
As a special case, for a Pauli-type generator with spectrum $\{\pm1/2\}$ one recovers the usual parameter-shift rule
\begin{align}
	F'(0)=\frac{F(\pi/2)-F(-\pi/2)}{2}.
\end{align}

\subsubsection*{Multi-parameter GPSR}

We now extend GPSR to the multi-parameter generating function
\begin{align}
	\mathcal G_{\mathbf A}(\vec\eta;t)=\mathcal G_{\mathbf A}(\eta_0,\dots,\eta_{L-1};t),\qquad
	\vec\eta=(\eta_0,\dots,\eta_{L-1}).
\end{align}
For each parameter $\eta_a$, the associated generator is $\mathbf B_a$ with finite discrete spectrum
\begin{align}
	\mathbf B_a=\sum_{s}\lambda_{a,s}P_{a,s},\qquad
	\Omega_a:=\{\lambda_{a,s}-\lambda_{a,s'}\}.
\end{align}
Choose shift points $\{s_{a,p}\}_{p=1}^{M_a}$ and solve the one-dimensional GPSR system
\begin{align}
	\sum_{p=1}^{M_a} c_{a,p}^{(r)} e^{i\omega s_{a,p}}=(i\omega)^r,\qquad \forall\,\omega\in\Omega_a,
	\label{eq:psr-1d-per-a}
\end{align}
which produces coefficients $c_{a,p}^{(r)}$ for each derivative order $r$.

For an order-$m$ term characterized by the multiplicities $\{\beta_a\}$ in Eq.~\eqref{eq:beta-definition}, the multi-parameter GPSR factorizes:
\begin{align}
	\left.
	\Bigg(\prod_{a=0}^{L-1}\frac{\partial^{\beta_a}}{\partial\eta_a^{\beta_a}}\Bigg)
	\mathcal G_{\mathbf A}(\vec\eta;t)
	\right|_{\vec\eta=\vec 0}
	=
	\sum_{p_0=1}^{M_0}\cdots\sum_{p_{L-1}=1}^{M_{L-1}}
	\Bigg[\prod_{a=0}^{L-1} c_{a,p_a}^{(\beta_a)}\Bigg]
	\mathcal G_{\mathbf A}\big(\eta_0=s_{0,p_0},\dots,\eta_{L-1}=s_{L-1,p_{L-1}};t\big).
	\label{eq:psr-multi-parameter}
\end{align}

Combining Eq.~\eqref{eq:psr-multi-parameter} with Eq.~\eqref{eq:chi-to-derivatives-channel} yields the GPSR estimator directly for the multi-pulse response coefficient in Eq.~\eqref{eq:delta-A-channel-delta}:
\begin{align}
	\boxed{
		\sum_{b_1=0}^{n_{a_1}-1}\cdots\sum_{b_m=0}^{n_{a_m}-1}
		\chi^{(m)}_{\mathbf A; a_1\cdots a_m}
		\big(t; t_{a_1b_1},\dots,t_{a_mb_m}\big)
		=
		\frac{1}{\prod_{a=0}^{L-1}\beta_a!}
		\sum_{p_0=1}^{M_0}\cdots\sum_{p_{L-1}=1}^{M_{L-1}}
		\Bigg[\prod_{a=0}^{L-1} c_{a,p_a}^{(\beta_a)}\Bigg]
		\mathcal G_{\mathbf A}\big(\eta_0=s_{0,p_0},\dots,\eta_{L-1}=s_{L-1,p_{L-1}};t\big).
	}
	\label{eq:chi-psr-unified}
\end{align}
In summary, the nonlinear response kernels can be expressed either (i) as Kubo-type nested commutators with explicit causality (Eqs.~\eqref{eq:delta-A(x)} and \eqref{eq:chi-kubo-causal}), or (ii) as derivatives of the finite-dimensional generating function at $\vec\eta=\vec 0$ under $\delta$-pulse drives.
GPSR turns these derivatives into a finite linear combination of shifted expectation values, yielding an exact and hardware-friendly route to estimating higher-order response functions from finite-shot measurements.

\section*{Appendix B: Complexity Analysis of the Generalized Parameter-Shift Rule (GPSR)}
\label{app:GPSR_complexity}

This appendix provides a self-contained resource analysis for the generalized parameter-shift rule (GPSR) used to reconstruct multi-time higher-order responses (equivalently, high-order pump derivatives). We decompose the overall cost into:
(i) the number of distinct circuit instances (shift configurations), and
(ii) the sampling cost (shots) required to achieve an additive precision $\epsilon$.

We consider an impulsive pump implemented by a unitary kick $e^{-i\eta \mathbf B}$ (single parameter) or a product of kicks $\prod_{a=0}^{L-1}e^{-i\eta_a \mathbf B_a}$ (multi-parameter). The experimentally measured observable is
\begin{equation}
	F(\eta)\equiv\langle \mathbf A(t)\rangle_\eta
	=
	\langle e^{i\eta \mathbf B}\mathbf A(t)e^{-i\eta \mathbf B}\rangle_0,
	\label{A1}
\end{equation}
or, in the multi-parameter setting,
\begin{equation}
	\mathcal G_{\mathbf A}(\vec\eta)\equiv\langle \mathbf A(t)\rangle_{\vec\eta},
	\qquad
	\vec\eta=(\eta_0,\dots,\eta_{L-1}).
	\label{A2}
\end{equation}
The goal of GPSR is to reconstruct the mixed derivative at $\vec\eta=\vec 0$ from finitely many shifted evaluations:
\begin{equation}
	\partial_{\vec\eta}^{\vec\beta}\mathcal G_{\mathbf A}(\vec 0),
	\qquad
	\vec\beta=(\beta_0,\dots,\beta_{L-1})\in\mathbb N^L,
	\label{A3}
\end{equation}
where $\vec\beta$ is a multi-index taking non-negative integers.

\subsection*{B.1  Single-parameter GPSR: shift count and exactness}

\subsubsection*{Finite spectrum $\Rightarrow$ finite Fourier support}

Assume that the generator $\mathbf B$ admits a finite discrete spectral decomposition,
\begin{equation}
	\mathbf B=\sum_{s\in S}\lambda_sP_s,
	\label{A4}
\end{equation}
where $P_s$ are projection operators satisfying $\sum_{s\in S}P_s=I$ and $P_sP_{s'}=\delta_{s,s'}P_s$. Then the measured expectation value $F(\eta)$ has the corresponding decomposition
\begin{equation}
	\begin{aligned}
		F(\eta)
		&=
		{\rm Tr}\!\left[\boldsymbol{\rho}_0\,e^{i\eta \mathbf B}\mathbf A(t)e^{-i\eta \mathbf B}\right] \\
		&=
		\sum_{s,s'\in S}e^{i\eta(\lambda_s-\lambda_{s'})}\,
		{\rm Tr}\!\left[\boldsymbol{\rho}_0P_s\mathbf A(t)P_{s'}\right].
	\end{aligned}
	\label{A5}
\end{equation}
Define the spectral-difference set
\begin{equation}
	\Omega=\{\lambda_s-\lambda_{s'}:s,s'\in S\},
	\qquad
	K\equiv|\Omega|,
	\label{A6}
\end{equation}
where $\Omega$ collects distinct frequencies. Here $\tilde F(\omega)$ denotes the exact Fourier coefficient in the finite Fourier expansion of $F(\eta)$, while $F_{\mathrm{est}}(\cdot)$ will be reserved for shot-noise estimators obtained from finite sampling.
Grouping identical frequencies yields a finite Fourier series
\begin{equation}
	F(\eta)=\sum_{\omega\in\Omega}\tilde F(\omega)e^{i\omega\eta},
	\qquad
	\tilde F(\omega)=\sum_{s,s':\lambda_s-\lambda_{s'}=\omega}{\rm Tr}\!\left[\boldsymbol{\rho}_0P_s\mathbf A(t)P_{s'}\right].
	\label{A7}
\end{equation}
Consequently, the $m$th derivative of $F(\eta)$ at $\eta=0$ is 
\begin{equation}
	F^{(m)}(0)
	=
	\sum_{\omega\in\Omega}\tilde F(\omega)(i\omega)^m .
	\label{A8}
\end{equation}

\subsubsection*{GPSR as a linear system: a lower bound on the number of shifts}

GPSR seeks shift points $\{s_p\}_{p=1}^M$ and coefficients $\{c_p^{(m)}\}_{p=1}^M$ such that
\begin{equation}
	\boxed{
		F^{(m)}(0)=\sum_{p=1}^M c_p^{(m)}F(s_p).
	}
	\label{eq:F-m-0-estimation}
\end{equation}
Substituting $F(s_p)=\sum_{\omega\in\Omega}\tilde F(\omega)e^{i\omega s_p}$ gives
\begin{equation}
	\sum_{p=1}^M c_p^{(m)}F(s_p)
	=
	\sum_{\omega\in\Omega} \tilde F(\omega)
	\left(\sum_{p=1}^M c_p^{(m)}e^{i\omega s_p}\right).
	\label{A10}
\end{equation}
For Eq.~(A10) to hold for arbitrary Fourier amplitudes $\{ \tilde F(\omega) \}$, one must satisfy, frequency by frequency,
\begin{equation}
	\boxed{
		\sum_{p=1}^M e^{i\omega s_p} c_p^{(m)}=(i\omega)^m,
		\qquad
		\forall\omega\in\Omega.
	}
	\label{A11}
\end{equation}
This linear system contains $K$ independent constraints for $M$ unknown coefficients. One can solve the coefficient $c_p^{(m)}$ classically if
\begin{equation}
	\boxed{
		M\ge K=|\Omega|.
	}
	\label{A12}
\end{equation}
In practice one often sets $M=K$ and chooses $\{s_p\}$ to improve the conditioning of the Vandermonde-type matrix $V_{\omega,p}=e^{i\omega s_p}$; the resulting coefficient norms directly control the sampling overhead.

\subsection*{B.2 Sampling complexity (shots) and coefficient-norm amplification}

After solving the coefficient $c_p^{(m)}$, the $m$th derivative $F^{(m)}(0)$ can be obtained according to Eq.~\eqref{eq:F-m-0-estimation} with $F(s_p)$ estimated on quantum computers. Assume $F(s_p)$ is estimated from $N_p$ shots by an unbiased estimator $F_{\mathrm{est}}(s_p)$.
The GPSR estimator is
\begin{equation}
	F^{(m)}_{\mathrm{est}}(0)=\sum_{p=1}^M c_p^{(m)}F_{\mathrm{est}}(s_p).
	\label{A13}
\end{equation}
Assuming independent sampling across shift points,
\begin{equation}
	{\rm Var}\!\left(F^{(m)}_{\mathrm{est}}(0)\right)
	=
	\sum_{p=1}^M (c_p^{(m)})^2\,{\rm Var}\!\left(F_{\mathrm{est}}(s_p)\right).
	\label{A14}
\end{equation}
For bounded observables (e.g., Pauli outcomes $\pm1$), ${\rm Var}\!\left(F_{\mathrm{est}}(s_p)\right)\le 1/N_p$.
Let $N_{\rm tot}=\sum_{p=1}^M N_p$ be the \emph{total} number of shots across all shifts.
Under uniform allocation $N_p=N_{\rm tot}/M$,
\begin{equation}
	{\rm Var}\!\left(F^{(m)}_{\mathrm{est}}(0)\right)
	\le
	\frac{M\,\|c^{(m)}\|_2^2}{N_{\rm tot}},
	\qquad
	\|c^{(m)}\|_2^2\equiv\sum_{p=1}^M(c_p^{(m)})^2.
	\label{A15}
\end{equation}
Therefore, to achieve an additive standard deviation $\epsilon$, it suffices to take
\begin{equation}
	\boxed{
		N_{\rm tot}\gtrsim\frac{M\,\|c^{(m)}\|_2^2}{\epsilon^2}.
	}
	\label{A16}
\end{equation}

While the bounds above are exact, they are worst-case with respect to the choice of the pump generator $\mathbf B$.
For a generic many-body operator $\mathbf B$ acting on a Hilbert space of dimension $d$ with an essentially non-degenerate spectrum, the spectral-difference set may reach
$|\Omega|=\Theta(d^2)$, implying that the number of required shifts can grow exponentially with system size.
Therefore, GPSR-based measurements are efficient only under physically motivated restrictions on the pump operator.

In this work we restrict to physically implementable impulsive pumps, where each kick generator is a low-weight local operator, or a sum of commuting local terms supported on a region of size $r$ (independent of, or at most polynomial in, the total system size $N$). A representative choice is
\begin{equation}
	\mathbf B=\sum_{j\in \mathcal S} b_j,\qquad [b_j,b_{j'}]=0,\qquad \|b_j\|=O(1),\qquad r\equiv|\mathcal S|.
\end{equation}
In this setting, the spectrum of $\mathbf B$ is confined to an interval of width $O(r)$, and the number of distinct gaps satisfies
\begin{equation}
	K=|\Omega|=O(r),
\end{equation}
so exact reconstruction requires only $M=O(r)$ shift points (and $M=O(1)$ when $r=O(1)$).

Moreover, choosing Fourier-grid shifts $s_p=2\pi(p-1)/M$ typically yields a well-conditioned reconstruction matrix (equivalently, a discrete Fourier transform over the gap set). One then obtains the rough scaling estimate
\begin{equation}
	\|c^{(m)}\|_2^2
	=
	O\!\left(\frac{1}{M}\sum_{\omega\in\Omega}|\omega|^{2m}\right)
	=
	O(r^{2m}),
\end{equation}
which leads, for fixed derivative order $m$, to a polynomial total-shot requirement
\begin{equation}
	N_{\rm tot}
	=
	O\!\left(\frac{M\|c^{(m)}\|_2^2}{\epsilon^2}\right)
	=
	O\!\left(\frac{r^{2m+1}}{\epsilon^2}\right).
\end{equation}

In our exemplary models, each impulsive pump is generated by a single-site or two-site Pauli operator (or a short sum of commuting local terms), hence $r=O(1)$ and $K=O(1)$ (e.g., eigenvalues $\pm1/2$ yield gaps $\Omega=\{2\}$). Accordingly, both the shift count $M$ and the coefficient amplification $\|c^{(m)}\|_2$ remain bounded as the system size increases, justifying the efficiency claim in $N$ at fixed response order.

This shows that the sampling cost is governed by both the shift count $M$ and the coefficient amplification $\|c^{(m)}\|_2$, which depends on the spectral structure $\Omega$ and the conditioning of the reconstruction linear system. In typical physical settings with local (or commuting) pumps supported on $r$ sites, one has $|\Omega|=O(r)$ and an exact GPSR reconstruction uses $M=O(r)$ shifts (in particular $M=O(1)$ when $r=O(1)$). For Fourier-grid shifts $s_p=2\pi(p-1)/M$ and fixed derivative order $m$, the coefficient norm scales as $\|c^{(m)}\|_2^2=O(r^{2m})$, leading to a polynomial shot complexity $N_{\rm tot}=O(r^{2m+1}/\epsilon^2)$. In our exemplary models, $\mathbf B$ is a one- or two-site Pauli operator (or a short commuting sum), hence $r=O(1)$ and both $M$ and $\|c^{(m)}\|_2$ remain bounded with system size.

\subsection*{B.3 Multi-parameter GPSR: circuit count and factorized error bounds}

We now consider $L$ pump parameters $\vec\eta=(\eta_0,\dots,\eta_{L-1})$ and a multi-index $\vec\beta=(\beta_0,\dots,\beta_{L-1})$. For each direction $a$, GPSR uses $M_a$ shifts $\{s_{a,p_a}\}_{p_a=1}^{M_a}$ and one-dimensional coefficients $\{c_{a,p_a}^{(\beta_a)}\}$.

\subsubsection*{Circuit instances: Cartesian product scaling}

GPSR evaluates $\mathcal G_{\mathbf A}(\vec\eta)$ on the Cartesian product grid
\begin{equation}
	\vec s_{\vec p}=(s_{0,p_0},\dots,s_{L-1,p_{L-1}}),
	\qquad
	\vec p=(p_0,\dots,p_{L-1}),
	\qquad
	p_a\in\{1,\dots,M_a\}.
	\label{A17}
\end{equation}
Hence the number of distinct circuit instances is
\begin{equation}
	\boxed{
		N_{\rm circ}=\prod_{a=0}^{L-1}M_a.
	}
	\label{A18}
\end{equation}
If $\beta_a=0$ (no derivative in direction $a$), one may set $M_a=1$, giving
\begin{equation}
	N_{\rm circ}=\prod_{a\in\mathcal A}M_a,
	\qquad
	\mathcal A\equiv\{a:\beta_a>0\}.
	\label{A19}
\end{equation}
Moreover, applying the single-parameter argument to each $\mathbf B_a$ with its spectral-difference set $\Omega_a$, exact reconstruction generically requires
\begin{equation}
	\boxed{
		M_a\ge|\Omega_a|
		\quad\Rightarrow\quad
		N_{\rm circ}\ge\prod_{a\in\mathcal A}|\Omega_a|.
	}
	\label{A20}
\end{equation}

\subsubsection*{Factorized variance bounds: uniform and optimal shots}

The multi-parameter GPSR estimator can be written as
\begin{equation}
	\Xi_{\mathrm{est}}
	=
	\sum_{\vec p}C_{\vec p}\,\mathcal G_{\mathbf A,\mathrm{est}}(\vec s_{\vec p}),
	\qquad
	C_{\vec p}=\prod_{a=0}^{L-1}c_{a,p_a}^{(\beta_a)}.
	\label{A21}
\end{equation}
Assuming independent sampling across configurations $\vec p$,
\begin{equation}
	{\rm Var}\!\left(\Xi_{\mathrm{est}}\right)
	=
	\sum_{\vec p}C_{\vec p}^2\,\frac{{\rm Var}_{\vec p}}{N_{\vec p}}.
	\label{A22}
\end{equation}

\paragraph{(i) Uniform shots.}
If ${\rm Var}_{\vec p}\le 1$ and $N_{\vec p}=N_{\rm shot}$ for all $\vec p$, then
\begin{equation}
	{\rm Var}\!\left(\Xi_{\mathrm{est}}\right)
	\le
	\frac{1}{N_{\rm shot}}\sum_{\vec p}C_{\vec p}^2.
	\label{A23}
\end{equation}
The product structure yields exact factorization,
\begin{equation}
	\sum_{\vec p}C_{\vec p}^2
	=
	\prod_{a=0}^{L-1}\left(\sum_{p_a=1}^{M_a}\left(c_{a,p_a}^{(\beta_a)}\right)^2\right)
	=
	\prod_{a=0}^{L-1}\|c_a^{(\beta_a)}\|_2^2.
	\label{A24}
\end{equation}
Thus a sufficient condition for precision $\epsilon$ is
\begin{equation}
	\boxed{
		N_{\rm shot}\gtrsim\frac{\prod_{a=0}^{L-1}\|c_a^{(\beta_a)}\|_2^2}{\epsilon^2}.
	}
	\label{A25}
\end{equation}
The resulting total shot cost under uniform allocation is
\begin{equation}
	\boxed{
		N_{\rm tot}=N_{\rm circ}\,N_{\rm shot}
		\gtrsim
		\frac{\left(\prod_{a\in\mathcal A}M_a\right)
			\left(\prod_{a\in\mathcal A}\|c_a^{(\beta_a)}\|_2^2\right)}{\epsilon^2}.
	}
	\label{A26}
\end{equation}

\paragraph{(ii) Optimal shots under a fixed budget.}

If one allows non-uniform $N_{\vec p}$ under a fixed total budget $N_{\rm tot}=\sum_{\vec p}N_{\vec p}$, the variance is minimized by allocating
\begin{equation}
	N_{\vec p}\propto|C_{\vec p}|\sqrt{{\rm Var}_{\vec p}}.
	\label{A27}
\end{equation}
Under the conservative bound ${\rm Var}_{\vec p}\le 1$, one obtains
\begin{equation}
	\sqrt{{\rm Var}_{\min}\!\left(\Xi_{\mathrm{est}}\right)}
	\lesssim
	\frac{\sum_{\vec p}|C_{\vec p}|}{\sqrt{N_{\rm tot}}}.
	\label{A28}
\end{equation}
Again, the $\ell_1$ weight factorizes,
\begin{equation}
	\sum_{\vec p}|C_{\vec p}|
	=
	\prod_{a=0}^{L-1}\left(\sum_{p_a=1}^{M_a}\left|c_{a,p_a}^{(\beta_a)}\right|\right)
	=
	\prod_{a=0}^{L-1}\|c_a^{(\beta_a)}\|_1.
	\label{A29}
\end{equation}
Therefore, a sufficient condition for precision $\epsilon$ is
\begin{equation}
	\boxed{
		N_{\rm tot}\gtrsim\frac{\left(\prod_{a=0}^{L-1}\|c_a^{(\beta_a)}\|_1\right)^2}{\epsilon^2}.
	}
	\label{A30}
\end{equation}
This bound is typically tighter than the uniform-allocation $\ell_2$ bound.

\subsection*{B.4 Summary}

The dominant GPSR resource scalings are controlled by:
(i) the spectral-difference support $|\Omega_a|$ (lower bounding $M_a$),
(ii) the Cartesian-product circuit count $N_{\rm circ}=\prod_{a\in\mathcal A}M_a$,
and (iii) coefficient-norm amplification that governs shot complexity. In particular,
\begin{equation}
	N_{\rm shot}\gtrsim\frac{\prod_a\|c_a^{(\beta_a)}\|_2^2}{\epsilon^2}
	\quad\text{(uniform allocation)},
	\qquad
	N_{\rm tot}\gtrsim\frac{\left(\prod_a\|c_a^{(\beta_a)}\|_1\right)^2}{\epsilon^2}
	\quad\text{(optimal allocation)}.
	\label{A31}
\end{equation}

Finally, we emphasize that efficiency in the system size holds for fixed response order
and physically implementable pumps (low-weight/local $\mathbf B_a$ or commuting sums on $O(1)$ support),
for which $|\Omega_a|$ and the coefficient norms remain at most polynomial in $N$.

\section*{Appendix C: Direct Construction of Time- and Frequency-Domain Response Functions}

For completeness, beyond the GPSR-based estimator and interpolation baseline, we also implemented a direct extraction of response functions; however, this approach is inferior to PSR and is only reliable when the perturbation strength $\eta$ is carefully chosen, which severely limits its applicability.

As a baseline method for response extraction, we directly computed the first-order response function $\chi^{(1)}(t)$ from experimental measurements of the observable $\langle \mathbf A(t)\rangle_{\eta}$ at various perturbation strengths $\eta$. The linear response was estimated by rescaling the signal at finite $\eta$ and subtracting the zero-field background. Specifically, we used
\begin{align}
	\chi^{(1)}(t) \approx \frac{\langle \mathbf A(t)\rangle_{s_1} - \langle \mathbf A(t)\rangle_{0}}{s_1},
\end{align}
for a small but finite $s_1$. Time-domain profiles for $\chi^{(1)}(t)$ at different $\eta$ values were compared against noiseless simulations.

\begin{figure}[H]
	\centering
	\includegraphics[width=1.0\textwidth]{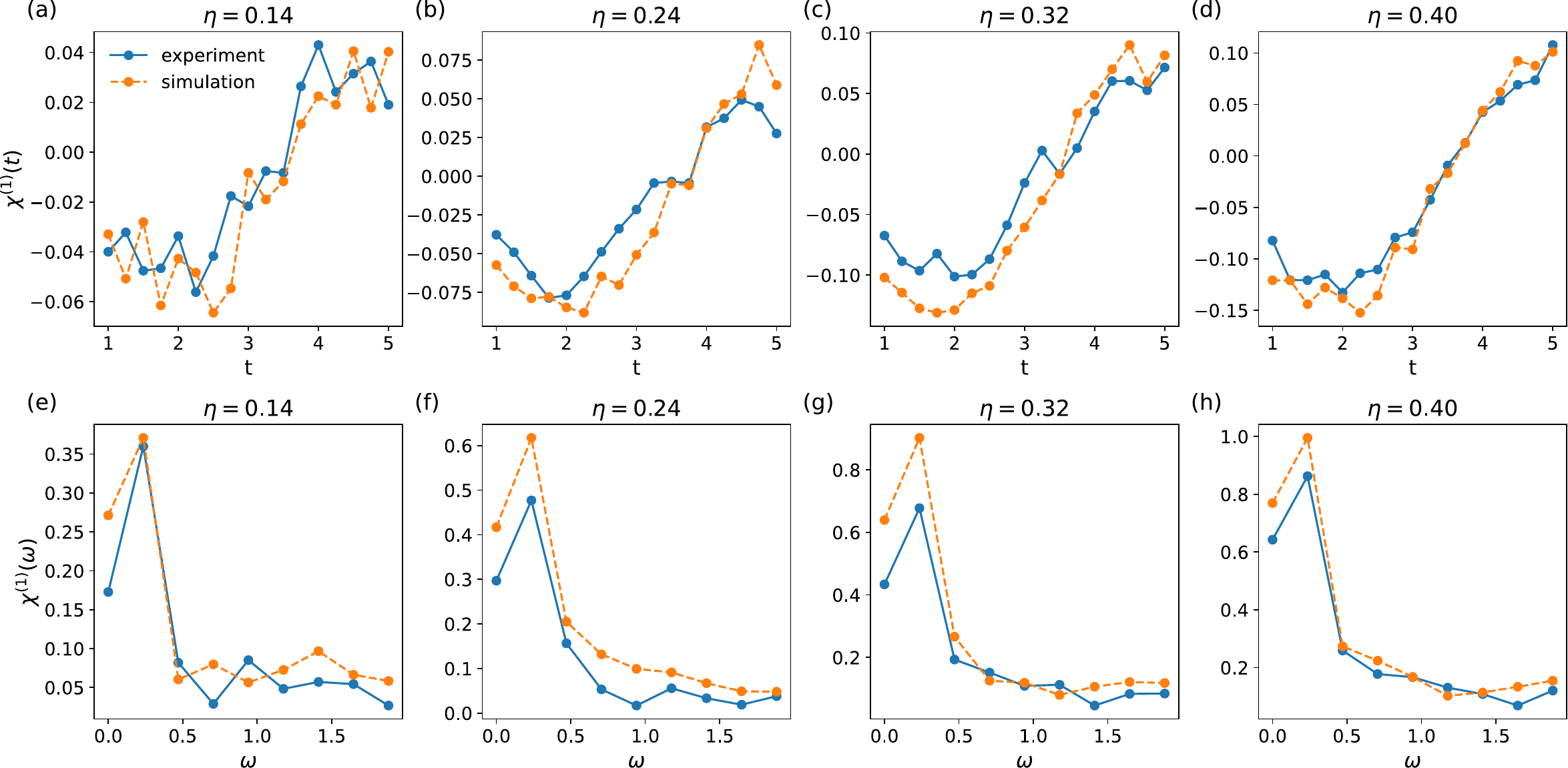}
	\caption{First-order response functions $\chi^{(1)}(t)$ and $\chi^{(1)}(\omega)$ at various perturbation strengths.
		Panels (a)--(d) show the time-domain signals at $\eta = 0.14$, 0.24, 0.32, and 0.40, respectively.
		Panels (e)--(h) present the corresponding frequency-domain spectra.
		Blue lines denote experimental results; orange dashed lines correspond to noiseless simulations.
		In the low $\eta$ regime (e.g., $\eta = 0.14$), larger fluctuations are observed due to lower signal-to-noise ratio, whereas the agreement improves significantly for larger $\eta$, illustrating the robustness of linear-response extraction at moderate perturbation strengths.}
	\label{a}
\end{figure}

Beyond the linear term, we also employed a stepwise signal decomposition in the spirit of standard response theory. We assume a power-series expansion of the pump-dependent signal,
\begin{align}
	\langle \mathbf A(t)\rangle_{\eta} \approx A^{0}(t) + \eta\,A^{1}(t) + \eta^3 A^{3}(t) + \eta^5 A^{5}(t) + \cdots,
\end{align}
where $A^{n}(t)$ collects the $n$-th order contribution to the response (with odd orders selected by symmetry in our setup). Given measurements at several pump amplitudes $s_1,s_2,s_3$ in a perturbative window, one can regard
\begin{align}
	\begin{pmatrix}
		\langle \mathbf A(t)\rangle_{s_1} - \langle \mathbf A(t)\rangle_{0} \\
		\langle \mathbf A(t)\rangle_{s_2} - \langle \mathbf A(t)\rangle_{0} \\
		\langle \mathbf A(t)\rangle_{s_3} - \langle \mathbf A(t)\rangle_{0}
	\end{pmatrix}
	\approx
	\begin{pmatrix}
		s_1 & s_1^3 & s_1^5 \\
		s_2 & s_2^3 & s_2^5 \\
		s_3 & s_3^3 & s_3^5
	\end{pmatrix}
	\begin{pmatrix}
		A^{1}(t) \\
		A^{3}(t) \\
		A^{5}(t)
	\end{pmatrix},
\end{align}
and solve this $3\times 3$ linear system for $A^{1}(t)$, $A^{3}(t)$, and $A^{5}(t)$. In practice, instead of explicitly inverting the matrix, we use analytically simplified combinations that successively cancel lower-order terms.

To isolate the third-order component, we define
\begin{align}
	\tilde{A}_3(t) = \langle \mathbf A(t)\rangle_{s_2} - \frac{s_2}{s_1}\langle \mathbf A(t)\rangle_{s_1},
\end{align}
which eliminates the leading linear term and leaves a dominant third-order signal proportional to $A^{3}(t)$, up to higher-order corrections. The corresponding third-order response function $\chi^{(3)}(\omega)$ is obtained by taking the discrete Fourier transform of $\tilde{A}_3(t)$ and applying a suitable normalization.

For the fifth-order term, we perform one more subtraction step:
\begin{align}
	\tilde{A}_5(t) = \langle \mathbf A(t)\rangle_{s_3} - \frac{s_3}{s_1}\langle \mathbf A(t)\rangle_{s_1} - \left( \frac{s_3^3}{s_2^3} \right) \left[ \langle \mathbf A(t)\rangle_{s_2} - \frac{s_2}{s_1}\langle \mathbf A(t)\rangle_{s_1} \right].
\end{align}
This expression cancels both the linear and third-order contributions under the polynomial ansatz, leaving a residual signal dominated by $s_3^5 A^{5}(t)$, from which the fifth-order response $A^{5}(t)$ and its Fourier transform $\chi^{(5)}(\omega)$ can be extracted.

\begin{figure}[H]
	\centering
	\includegraphics[width=1.0\textwidth]{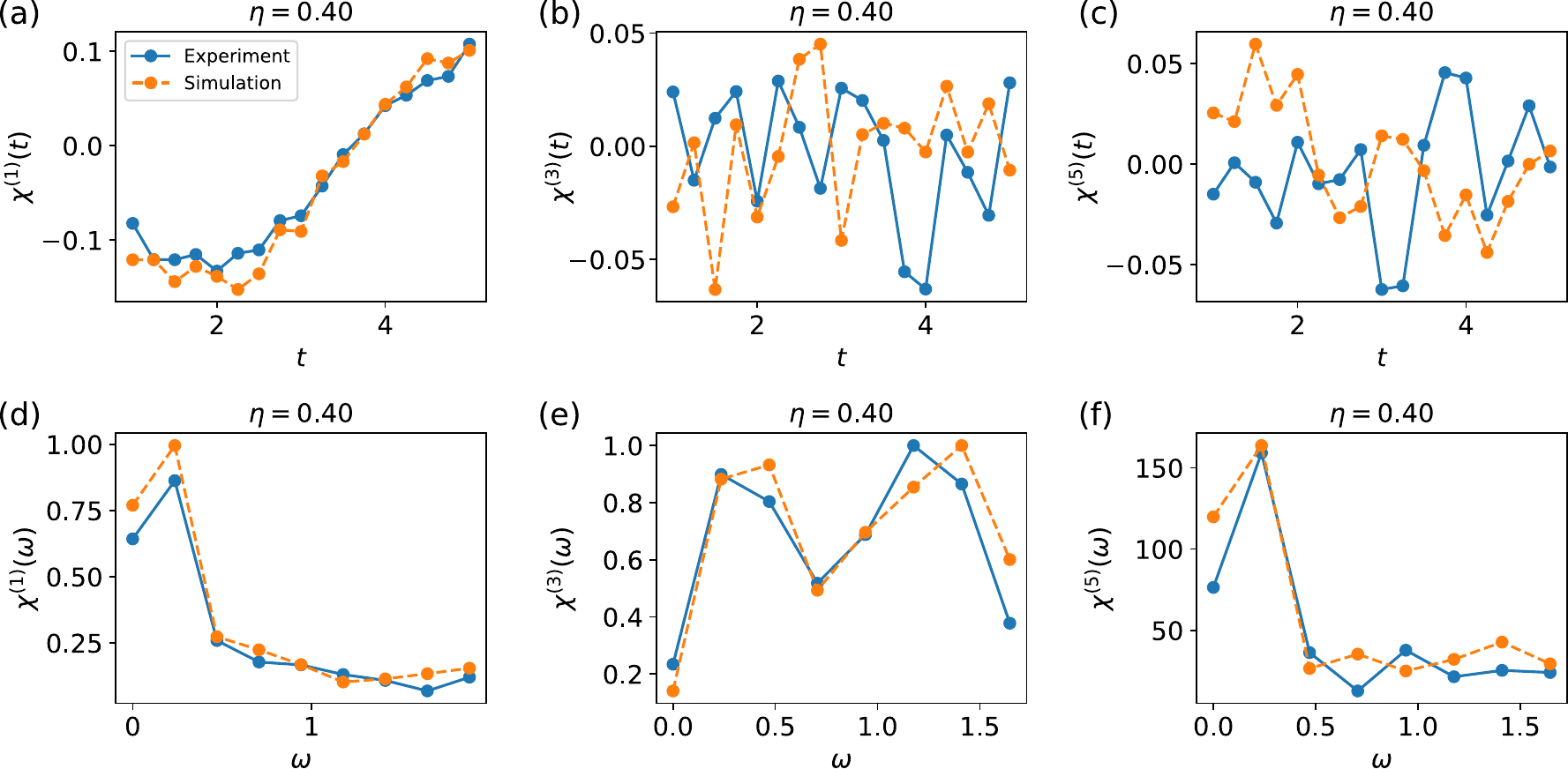}
	\caption{
		Time- and frequency-domain nonlinear response functions at $\eta = 0.40$, computed using the direct subtraction method.
		(a)--(c) show the time-domain responses $\chi^{(1)}(t)$, $\chi^{(3)}(t)$, and $\chi^{(5)}(t)$, obtained by successively removing lower-order contributions from raw signals.
		(d)--(f) present the corresponding frequency-domain spectra $\chi^{(1)}(\omega)$, $\chi^{(3)}(\omega)$, and $\chi^{(5)}(\omega)$ via discrete Fourier transform.
		Solid blue lines denote experimental results, and dashed orange lines represent noiseless simulations.
		While the main low-frequency structures are captured, higher-order components suffer from visible fluctuations, especially in the time domain.}
	\label{b}
\end{figure}

The extracted responses $\chi^{(1)}(t)$, $\chi^{(3)}(t)$, and $\chi^{(5)}(t)$, as well as their corresponding Fourier spectra, are shown in Fig.~\ref{b} for the representative case of $\eta = 0.40$. These results demonstrate how higher-order contributions and the associated nonlinear spectra can be resolved by combining measurements at multiple pump strengths under a controlled power-series expansion.

Overall, the direct finite-difference and stepwise subtraction methods provide a conceptually simple, response-theory--style route to constructing time- and frequency-domain response functions. However, their performance is limited by the choice of $\eta$ and by noise amplification at higher orders.

\section*{Appendix D: Response Function of Magnetization}

In the study of quantum many-body systems, the evaluation of magnetization plays a central role in probing the dynamic response of a spin system to external perturbations. This work focuses on the calculation of magnetization in the $x$-direction, denoted $\langle M^x(t) \rangle$, within a finite quantum spin chain governed by an XXZ Hamiltonian with a static, spatially modulated transverse field.

The unperturbed Hamiltonian of the system is defined as
\begin{align}
	\mathbf H_0 = \frac{1}{4}\sum_{j=0}^{N-2} \left( X_j X_{j+1} + Y_j Y_{j+1} + \Delta Z_j Z_{j+1} \right) - \frac{h_e}{2} \sum_{j=0}^{N-1} Z_j,
\end{align}
where $X_j,Y_j,Z_j$ are Pauli operators acting on site $j$, $\Delta$ is the anisotropy parameter, $h_e$ is the longitudinal magnetic field strength, and $N$ is the total number of spins. A transverse perturbation is introduced in the form
\begin{align}
	\mathbf B = \sum_{j=0}^{N-1} \cos\left( \frac{2\pi j}{N} \right) X_j,
\end{align}
and the full Hamiltonian becomes
\begin{align}
	\mathbf H = \mathbf H_0 + \eta \mathbf B,
\end{align}
where $\eta$ denotes the perturbation strength.

The observable of interest is the transverse magnetization per site, defined as
\begin{align}
	\mathbf M^x = -\frac{1}{N} \sum_{j=0}^{N-1} X_j.
\end{align}
This operator captures the average spin alignment along the $x$-axis. The ground state $|\psi_0\rangle$ of the full Hamiltonian $\mathbf H$ is obtained through exact diagonalization, and the time-evolved state is given by
\begin{align}
	|\psi(t)\rangle = U(t) |\psi_0\rangle, \quad U(t) = e^{-i\mathbf H t}.
\end{align}
The time-dependent magnetization is then computed as the quantum expectation value
\begin{align}
	\langle M^x(t) \rangle = \langle \psi(t) | \mathbf M^x | \psi(t) \rangle.
\end{align}
Numerically, this is evaluated by
\begin{align}
	\langle M^x(t) \rangle = \operatorname{Re} \left[ \psi(t)^\dagger \mathbf M^x \psi(t) \right],
\end{align}
where $\psi(t) \in \mathbb{C}^{2^N}$ and $\mathbf M^x$ is the corresponding Hermitian matrix operator.

\begin{figure}[H]
	\centering
	\includegraphics[width=0.8\textwidth]{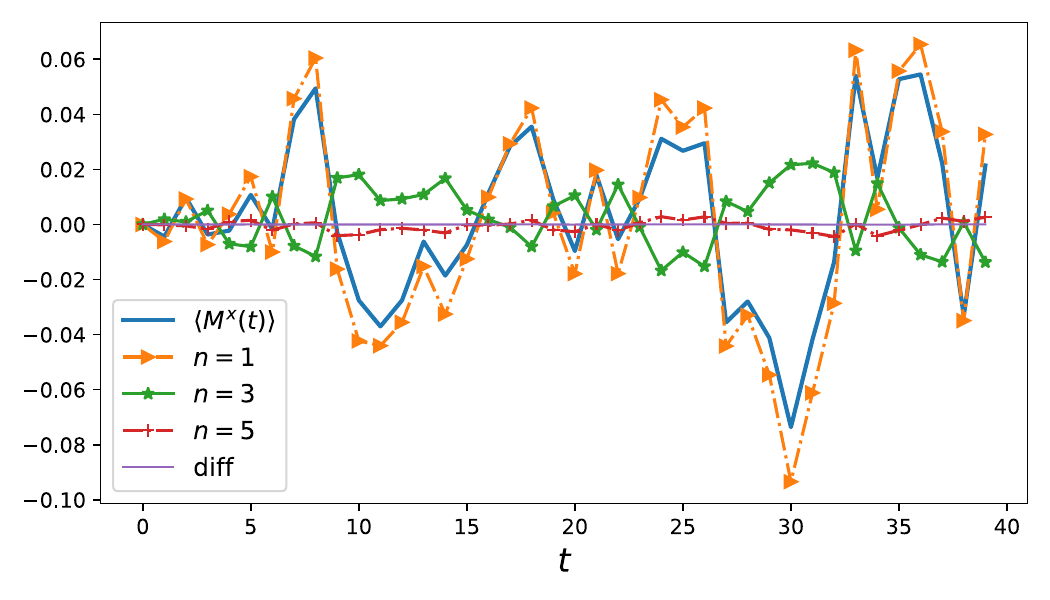}
	\caption{Magnetization signal $\langle M^x(t)\rangle_\eta$ and its extracted response contributions in the $\eta$ expansion. The $n=0$ term corresponds to the unpumped baseline $\langle M^x(t)\rangle_{\eta=0}$. Due to symmetry, the even-order contributions (e.g., $n=2,4$) vanish within numerical precision, and we display the leading nonzero odd orders ($n=1,3,5$). The curve labeled ``diff'' denotes the residual between the measured $\langle M^x(t)\rangle_\eta$ and the reconstructed partial sum obtained by adding the retained (nonzero) response contributions together with the $n=0$ baseline, serving as a consistency check.}
	\label{6}
\end{figure}

\begin{figure}[H]
	\centering
	\includegraphics[width=0.8\textwidth]{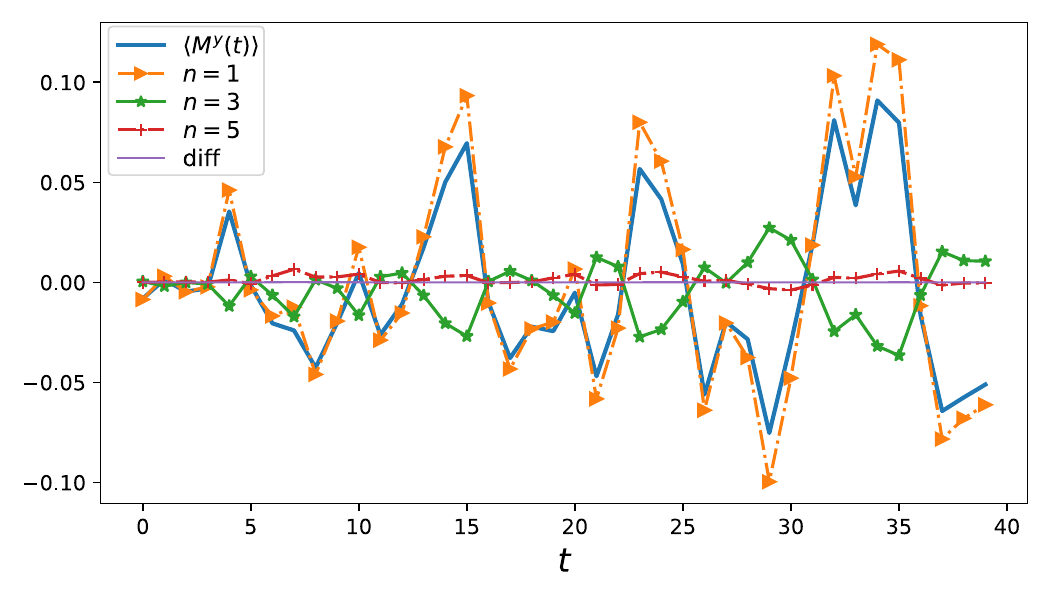}
	\caption{Magnetization signal $\langle M^y(t)\rangle_\eta$ and its extracted response contributions in the $\eta$ expansion. The $n=0$ term corresponds to the unpumped baseline $\langle M^y(t)\rangle_{\eta=0}$. Due to symmetry, the even-order contributions (e.g., $n=2,4$) vanish within numerical precision, and we display the leading nonzero odd orders ($n=1,3,5$). The curve labeled ``diff'' denotes the residual between the measured $\langle M^y(t)\rangle_\eta$ and the reconstructed partial sum obtained by adding the retained (nonzero) response contributions together with the $n=0$ baseline, serving as a consistency check.}
	\label{7}
\end{figure}

To study nonlinear effects, the magnetization is expanded as a power series in $\eta$ using nested commutators. Assuming the dynamics are governed by the unperturbed evolution operator $U_0(t) = e^{-i\mathbf H_0 t}$ and the observable in the Heisenberg picture is $\mathbf M^x(t) = U_0^\dagger(t) \mathbf M^x U_0(t)$, the expansion takes the form
\begin{align}
	\langle \mathbf M^x(t) \rangle_\eta 
	&= \langle \mathbf M^x(t) \rangle_0 
	+ i\eta \langle [\mathbf B, \mathbf M^x(t)] \rangle_0 
	+ \frac{(i\eta)^2}{2!} \langle [\mathbf B, [\mathbf B, \mathbf M^x(t)]] \rangle_0 \nonumber \\
	&\quad + \frac{(i\eta)^3}{3!} \langle [\mathbf B, [\mathbf B, [\mathbf B, \mathbf M^x(t)]]] \rangle_0 + \cdots,
\end{align}
where the expectation values are taken with respect to the ground state of $\mathbf H_0$. In practice, this expansion is carried out up to the seventh order to isolate linear and higher-order nonlinear contributions. The framework provides a comprehensive way to characterize dynamical response functions and can be extended to other forms of quantum perturbation theory. Illustrative results for the magnetization dynamics and the associated higher-order contributions are presented in Fig.~\ref{6} to Fig.~\ref{7}.

\section*{Appendix E: Response Function of Spin-Spin Correlation Function}

To characterize the spatial and dynamical correlations in the quantum spin chain, we compute the two-point and four-point spin correlation functions in various directions. The two-point spin correlation function between sites $i$ and $j$ along directions $\alpha$ and $\beta$ is defined as
\begin{align}
	C_{i,j}^{\alpha\beta}(t) = \langle \psi(t) | P_i^\alpha P_j^\beta | \psi(t) \rangle,
\end{align}
where $P_i^\alpha$ and $P_j^\beta$ are Pauli operators acting on sites $i$ and $j$ in directions $\alpha, \beta \in \{x, y, z\}$ respectively, and $|\psi(t)\rangle$ denotes the time-evolved state of the system under the full Hamiltonian $\mathbf H$. These quantities reflect the quantum correlations between different spin components and are sensitive to the anisotropy and external field parameters of the system. For instance, the $C_{0,1}^{xx}(t)$ correlation quantifies the alignment tendency between the $x$ components of neighboring spins at sites $0$ and $1$, and its temporal evolution provides insight into the propagation of local perturbations and collective excitations.

The evaluation is implemented numerically by representing the spin operators as tensor products of Pauli matrices and identity matrices in the full $2^N$-dimensional Hilbert space. Given the state vector $\psi(t)$ at time $t$, the correlation is computed as
\begin{align}
	C_{i,j}^{\alpha\beta}(t) = \operatorname{Re} \left[ \psi(t)^\dagger (P_i^\alpha P_j^\beta) \psi(t) \right],
\end{align}
where the operator $P_i^\alpha P_j^\beta$ is constructed via Kronecker products with appropriate identities acting on unaffected qubits. This method allows straightforward generalization to arbitrary spin directions and distances.

In addition to two-point correlations, higher-order correlations are also calculated to probe more complex quantum interactions and entanglement patterns. Specifically, we consider the four-point spin correlation function of the form
\begin{align}
	C_{i,j,k,l}^{\alpha\beta\gamma\delta}(t) = \langle \psi(t) | P_i^\alpha P_j^\beta P_k^\gamma P_l^\delta | \psi(t) \rangle,
\end{align}
with $\alpha, \beta, \gamma, \delta \in \{x, y, z\}$. As a representative example, the correlation $C_{0,1,2,3}^{xyxz}(t)$ involves spin operators along different directions and spatial locations, capturing intricate multi-body quantum correlations that are not accessible via two-point functions alone.

These multi-spin correlators are particularly important for identifying quantum phase transitions, topological order, or signatures of many-body localization. Their implementation follows the same principle: constructing the tensor product of four Pauli operators at the specified sites and evaluating the real part of the expectation value with the time-evolved state. The resulting data, resolved in time, reveals the interplay between spin components and the emergence of collective behavior under unitary dynamics.

\begin{figure}[H]
	\centering
	\includegraphics[width=0.8\textwidth]{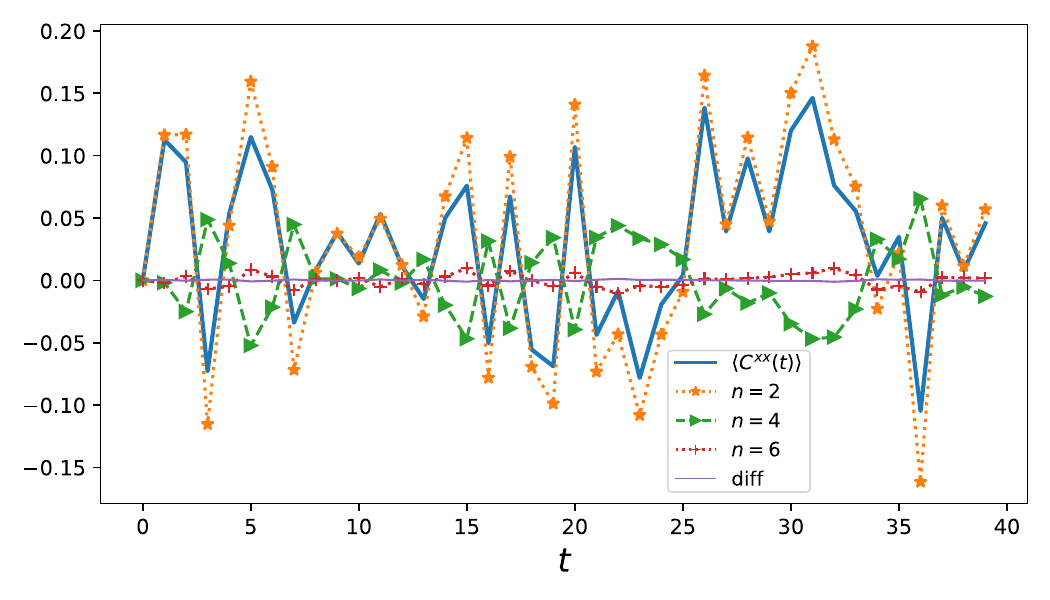}
	\caption{Two-point correlation signal $\langle C^{xx}(t)\rangle_\eta$ and its extracted response contributions in the $\eta$ expansion. The $n=0$ term corresponds to the unpumped baseline $\langle C^{xx}(t)\rangle_{\eta=0}$. Due to symmetry, the odd-order contributions vanish within numerical precision, and we display the leading nonzero even orders ($n=2,4,6$). The curve labeled ``diff'' denotes the residual between the measured $\langle C^{xx}(t)\rangle_\eta$ and the reconstructed partial sum obtained by adding the retained (nonzero) response contributions together with the $n=0$ baseline, serving as a consistency check of the decomposition.}
	\label{8}
\end{figure}

This framework provides a versatile and numerically exact method for exploring both linear and nonlinear quantum correlations in finite spin systems and lays the foundation for studying dynamical structure factors, response functions, and entanglement growth in driven quantum systems. Detailed results for the two-point and four-point spin correlation functions, along with their higher-order effects, are shown in Fig.~\ref{8} $\sim$ Fig.~\ref{10}.

\begin{figure}[H]
	\centering
	\includegraphics[width=0.8\textwidth]{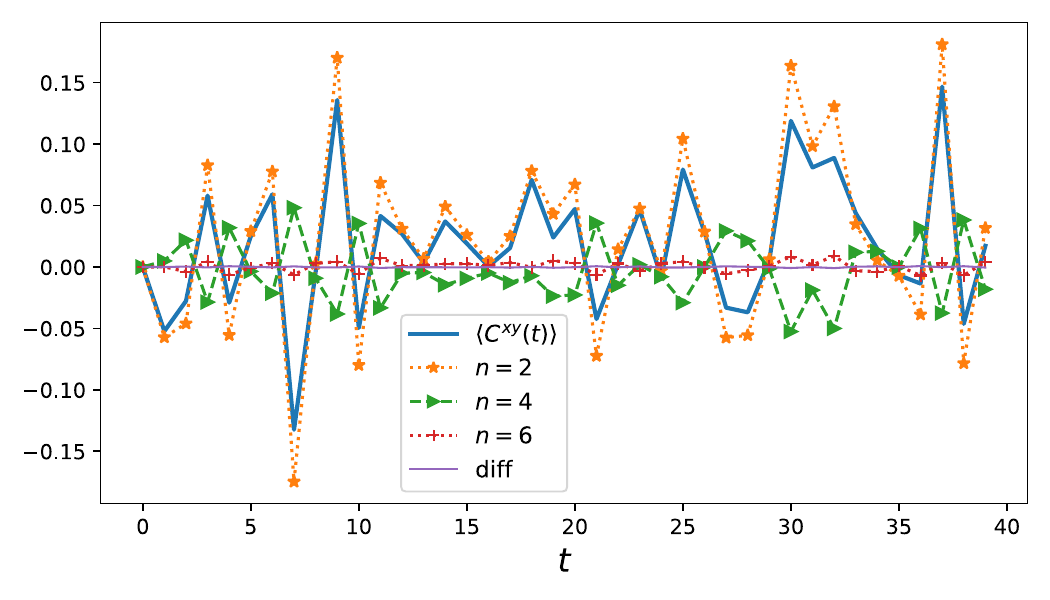}
	\caption{Two-point correlation signal $\langle C^{xy}(t)\rangle_\eta$ and its extracted response contributions in the $\eta$ expansion. The $n=0$ term corresponds to the unpumped baseline $\langle C^{xy}(t)\rangle_{\eta=0}$. Due to symmetry, the odd-order contributions vanish within numerical precision, and we display the leading nonzero even orders ($n=2,4,6$). The curve labeled ``diff'' denotes the residual between the measured $\langle C^{xy}(t)\rangle_\eta$ and the reconstructed partial sum obtained by adding the retained (nonzero) response contributions together with the $n=0$ baseline, validating the fidelity of the decomposition.}
	\label{9}
\end{figure}

\begin{figure}[H]
	\centering
	\includegraphics[width=0.8\textwidth]{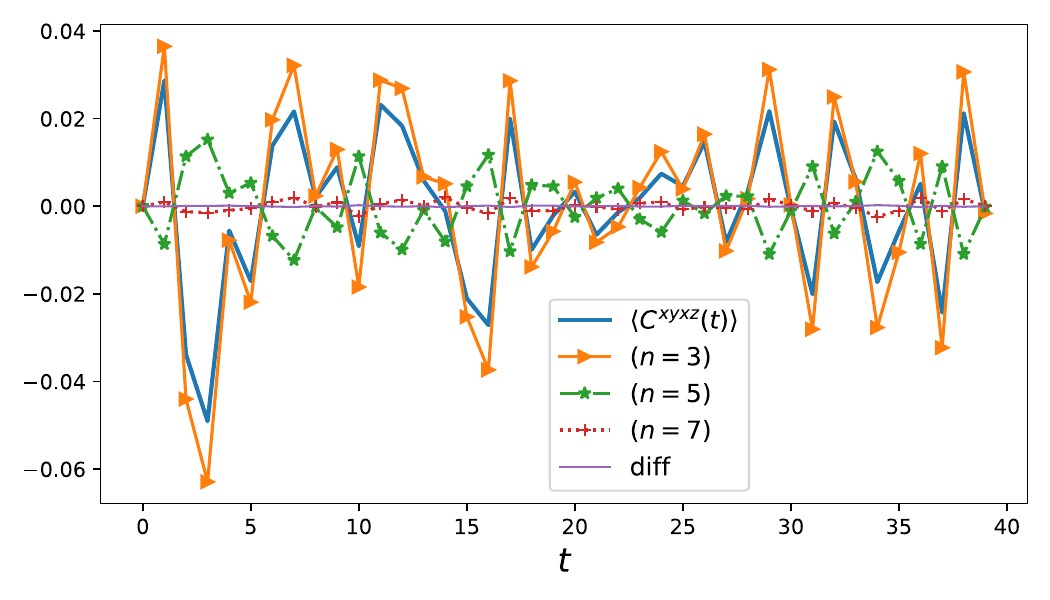}
	\caption{Four-point correlation signal $\langle C^{xyxz}(t)\rangle_\eta$ and its extracted response contributions in the $\eta$ expansion. The $n=0$ term corresponds to the unpumped baseline $\langle C^{xyxz}(t)\rangle_{\eta=0}$. Due to symmetry, the even-order contributions vanish within numerical precision, and we display the leading nonzero odd orders ($n=3,5,7$). The curve labeled ``diff'' denotes the residual between the measured $\langle C^{xyxz}(t)\rangle_\eta$ and the reconstructed partial sum obtained by adding the retained (nonzero) response contributions together with the $n=0$ baseline, confirming the correctness of the computed decomposition.}
	\label{10}
\end{figure}

\section*{Appendix F: Response Function of Current Density}

To investigate the dynamical transport properties of the spin chain, we evaluate the spin current operators along various directions. In quantum spin systems, the spin current between two adjacent sites $i$ and $j$ in direction $a \in \{x, y, z\}$ is generally defined through the continuity equation for spin density and takes the form of an antisymmetric combination of local spin operators. Specifically, the spin current operator along the $x$-direction is given by
\begin{align}
	J_{s,ij}^x = Y_i Z_j - Z_i Y_j,
\end{align}
where $X_i,Y_i,Z_i$ denotes the Pauli operator acting on site $i$ in the corresponding direction. Similarly, the spin current in the $y$- and $z$-directions is constructed by cyclic permutation of the Pauli matrices:
\begin{align}
	J_{s,ij}^y &= Z_i X_j - X_i Z_j, \\
	J_{s,ij}^z &= X_i Y_j - Y_i X_j.
\end{align}
These operators represent the flow of spin angular momentum between neighboring sites and are key observables in characterizing transport, dissipation, and symmetry breaking in spin systems.

In our numerical implementation, the spin current operator is encoded as a $2^N \times 2^N$ Hermitian matrix constructed via Kronecker products of the Pauli matrices and identity operators, such that the relevant operators act non-trivially only on sites $i$ and $j$. For a given time $t$, the quantum state of the system is evolved under the full Hamiltonian $\mathbf H$ from an initial ground state $|\psi_0\rangle$, resulting in
\begin{align}
	|\psi(t)\rangle = e^{-i\mathbf H t} |\psi_0\rangle.
\end{align}
The time-dependent expectation value of the spin current is then computed as
\begin{align}
	\langle J_{s,ij}^a(t) \rangle = \langle \psi(t) | J_{s,ij}^a | \psi(t) \rangle,
\end{align}
and in practice is evaluated numerically as
\begin{align}
	\langle J_{s,ij}^a(t) \rangle = \operatorname{Re} \left[ \psi(t)^\dagger J_{s,ij}^a \psi(t) \right].
\end{align}
In order to characterize the overall strength of the spin transport, we report the absolute value of this expectation value:
\begin{align}
	| \langle J_{s,ij}^a(t) \rangle |.
\end{align}
This measure reflects the instantaneous magnitude of spin flow, independent of its direction.

To further resolve the nonlinear contributions of spin transport under a modulated transverse field, we adopt a commutator-based expansion of the expectation value with respect to the perturbation strength $\eta$. For the unperturbed Hamiltonian $\mathbf H_0$ and corresponding time evolution operator $U_0(t) = e^{-i\mathbf H_0 t}$, the observable in the Heisenberg picture is
\begin{align}
	J_{s,ij}^a(t) = U_0^\dagger(t) J_{s,ij}^a U_0(t),
\end{align}
and its perturbative expansion in $\eta$ takes the form
\begin{align}
	\langle J_{s,ij}^a(t) \rangle_\eta 
	&= \langle J_{s,ij}^a(t) \rangle_0 
	+ i\eta \langle [\mathbf B, J_{s,ij}^a(t)] \rangle_0 
	+ \frac{(i\eta)^2}{2!} \langle [\mathbf B, [\mathbf B, J_{s,ij}^a(t)]] \rangle_0 \nonumber \\
	&\quad + \frac{(i\eta)^3}{3!} \langle [\mathbf B, [\mathbf B, [\mathbf B, J_{s,ij}^a(t)]]] \rangle_0 + \cdots,
\end{align}

where $\mathbf B$ is the transverse perturbation operator and the expectation values are taken with respect to the ground state of $\mathbf H_0$. This expansion allows for the decomposition of the observed current into linear and higher-order nonlinear contributions, enabling the identification of resonant response, symmetry-induced cancellations, and nontrivial transport dynamics beyond linear response.

\begin{figure}[H]
	\centering
	\includegraphics[width=0.93\textwidth]{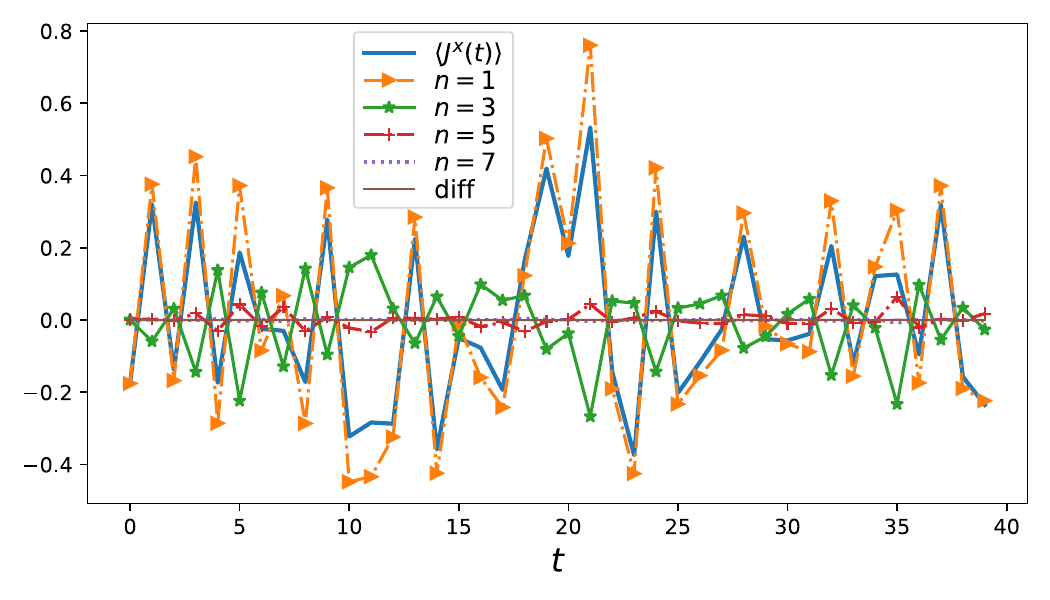}
	\caption{Current-density signal $\langle J^{x}(t)\rangle_\eta$ and its extracted response contributions in the $\eta$ expansion. The $n=0$ term corresponds to the unpumped baseline $\langle J^{x}(t)\rangle_{\eta=0}$. Due to symmetry, the even-order contributions vanish within numerical precision, and we display the leading nonzero odd orders ($n=1,3,5,7$). The curve labeled ``diff'' denotes the residual between the measured $\langle J^{x}(t)\rangle_\eta$ and the reconstructed partial sum obtained by adding the retained (nonzero) response contributions together with the $n=0$ baseline, serving as a consistency check.}
	\label{11}
\end{figure}

The resulting spin current data, resolved in both time and perturbation order, serve as a diagnostic of the system's dynamical behavior under driving, and provide a route to extracting transport coefficients and collective phenomena in strongly interacting spin models. Detailed results for the spin current, along with their higher-order effects, are shown in Fig.~\ref{11} $\sim$ Fig.~\ref{13}.

\begin{figure}[H]
	\centering
	\includegraphics[width=0.93\textwidth]{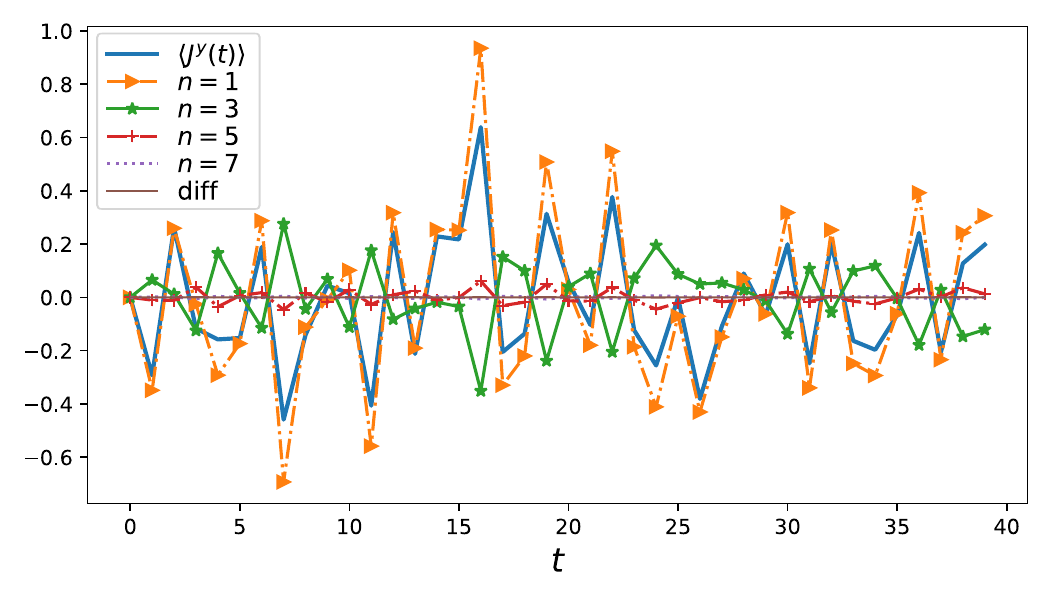}
	\caption{Current-density signal $\langle J^{y}(t)\rangle_\eta$ and its extracted response contributions in the $\eta$ expansion. The $n=0$ term corresponds to the unpumped baseline $\langle J^{y}(t)\rangle_{\eta=0}$. Due to symmetry, the even-order contributions vanish within numerical precision, and we display the leading nonzero odd orders ($n=1,3,5,7$). The curve labeled ``diff'' denotes the residual between the measured $\langle J^{y}(t)\rangle_\eta$ and the reconstructed partial sum obtained by adding the retained (nonzero) response contributions together with the $n=0$ baseline, serving as a consistency check.}
	\label{12}
\end{figure}

\begin{figure}[H]
	\centering
	\includegraphics[width=0.93\textwidth]{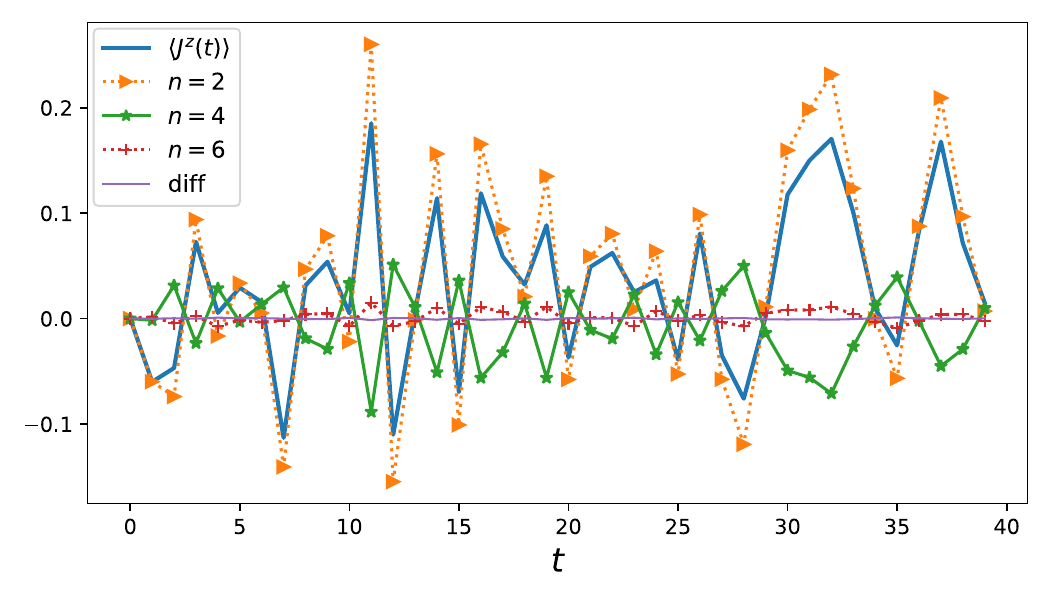}
	\caption{Current-density signal $\langle J^{z}(t)\rangle_\eta$ and its extracted response contributions in the $\eta$ expansion. The $n=0$ term corresponds to the unpumped baseline $\langle J^{z}(t)\rangle_{\eta=0}$. Due to symmetry, the odd-order contributions vanish within numerical precision, and we display the leading nonzero even orders ($n=2,4,6$). The curve labeled ``diff'' denotes the residual between the measured $\langle J^{z}(t)\rangle_\eta$ and the reconstructed partial sum obtained by adding the retained (nonzero) response contributions together with the $n=0$ baseline, serving as a consistency check.}
	\label{13}
\end{figure}

\section*{Appendix G: Scheme for Detecting quasi-particle Excitations via Response Function}

In addition to the mixed-field operator
\begin{align}
	\mathbf B = \sum_{i=0}^{N-1} (X_i + Z_i),
\end{align}
we also explored alternative forms of the perturbation operator $\mathbf B$ across different models. In the XXZ spin chain, we considered a purely transverse field perturbation,
\begin{align}
	\mathbf B = \sum_{i=0}^{N-1} X_i,
	\label{eq_xxz}
\end{align}
which still yields a well-defined quasi-particle spectrum with identifiable linear response contributions. 

For the Kitaev honeycomb model, the operator
\begin{align}
	\mathbf B = \sum_{i=0}^{N-1} Y_i
	\label{eq_honey}
\end{align}
was used to probe spin dynamics along the $y$-direction, which is naturally aligned with one of the anisotropic bond interactions in the model. Despite the model's topological nature, we observed that this form of $\mathbf B$ also produces clear linear response features in the time-domain signals.

In the case of the 2D toric code (Kitaev toric model), we adopted a generalized operator acting on a set of plaquette sites:
\begin{align}
	\mathbf B = \sum_{i \in \text{sites}} Y_i,
	\label{eq_toric}
\end{align}
which targets the creation and propagation of localized flux excitations. Remarkably, all these choices of $\mathbf B$ lead to nontrivial linear response behavior, suggesting that the linear components of the quasi-particle spectrum are robust features across various models and operator types. This highlights the universality of time-domain response techniques in capturing spectral signatures.

\begin{figure}[H]
	\centering
	\includegraphics[width=1.0\textwidth]{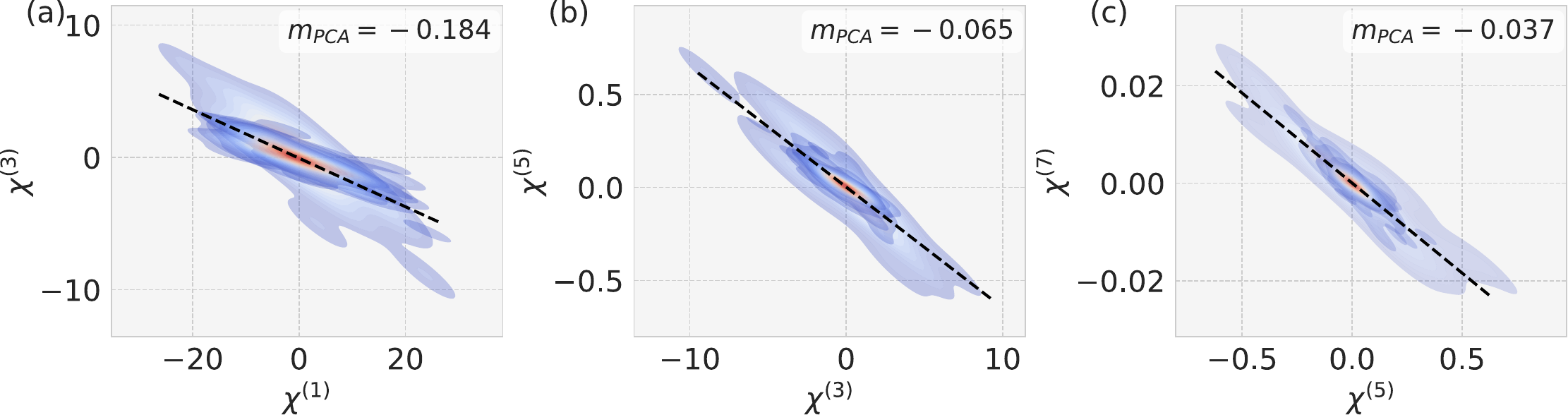}
	\caption{Excitation spectrum of the XXZ model under a purely transverse field perturbation (Eq.~\ref{eq_xxz}). The corresponding response function exhibits a clear linear trend, with $m_{\mathrm{PCA}}$ denoting the fitted slope obtained via principal component analysis.}
	\label{14}
\end{figure}
\begin{figure}[H]
	\centering
	\includegraphics[width=1.0\textwidth]{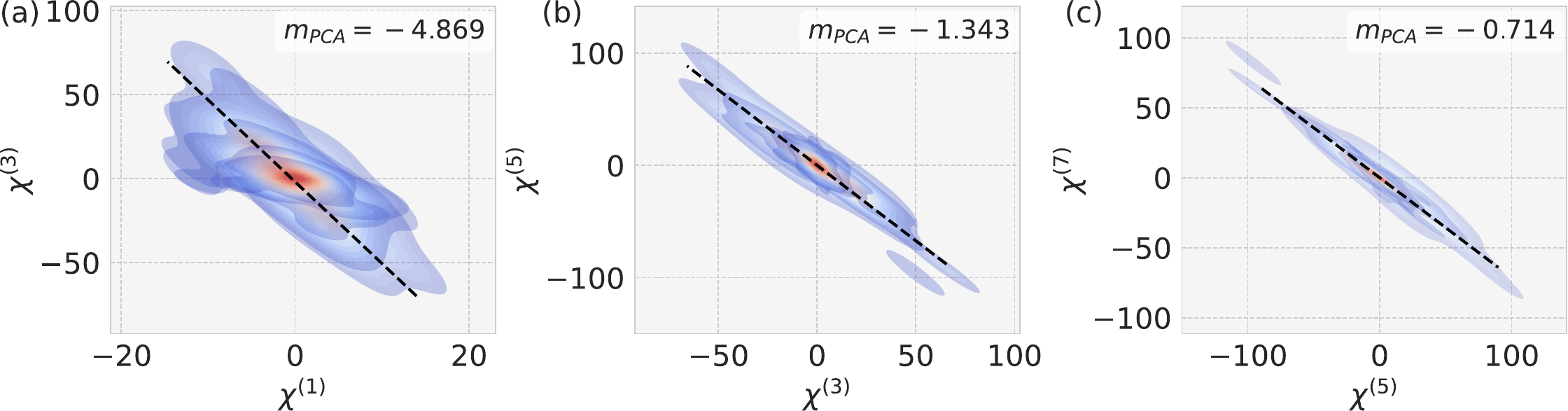}
	\caption{Excitation spectrum of the Kitaev honeycomb model subject to a purely transverse field perturbation (Eq.~\ref{eq_honey}). The response function maintains a well-defined linear dependence, and $m_{\mathrm{PCA}}$ represents the slope extracted through principal component fitting.}
	\label{15}
\end{figure}
\begin{figure}[H]
	\centering
	\includegraphics[width=1.0\textwidth]{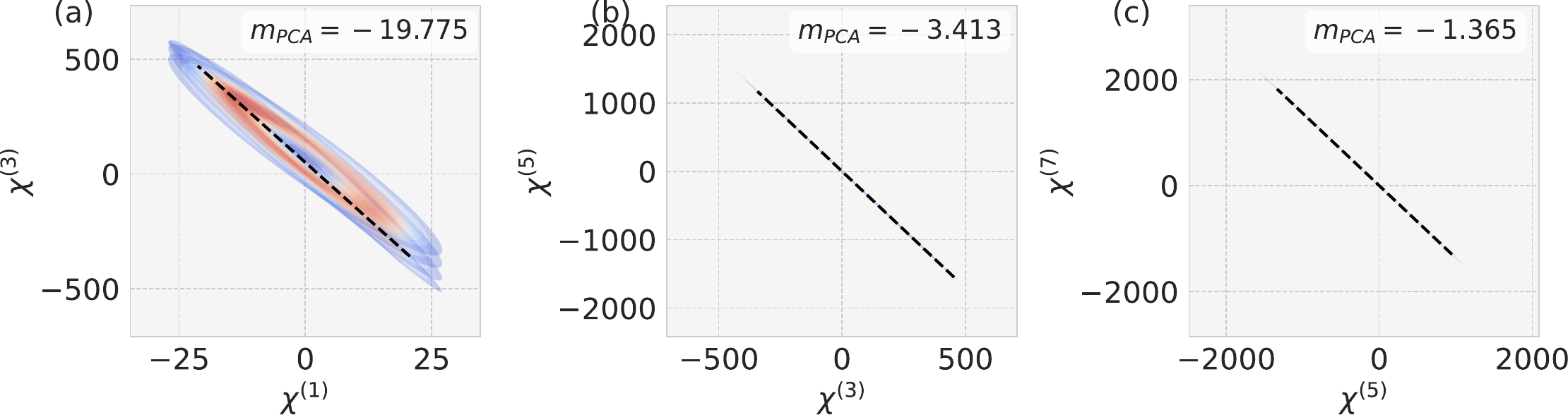}
	\caption{Excitation spectrum of the Kitaev toric code model under the application of a purely transverse field perturbation (Eq.~\ref{eq_toric}). The response function displays a pronounced linear relationship, with $m_{\mathrm{PCA}}$ indicating the slope obtained from principal component analysis.}
	\label{16}
\end{figure}

Future work will focus on a systematic comparison of quasi-particle excitation spectra across various models, parameter regimes, and distinct quantum phases to uncover model-dependent behaviors and phase-specific spectral signatures. In summary, this appendix collects the minimal technical details behind our response-extraction pipeline. We show how multi-time, higher-order response coefficients can be written as derivatives of a pulse-amplitude generating function and then evaluated exactly from a finite set of shifted circuits via the GPSR. Different observables can obey different symmetry selection rules, so the set of nonzero response orders may vary across the plots, while the “diff” curve serves as a practical consistency check of the reconstruction. Looking ahead, it would be useful to optimize the choice of shift points and measurement budget to further improve the stability of higher-order components under finite-shot noise.

\end{document}